\documentclass[11pt]{article}
\def\be{\begin{equation}}
\def\ee{\end{equation}}
\def\bea{\begin{eqnarray}}
\def\eea{\end{eqnarray}}
\usepackage{graphicx} 
\usepackage{amssymb,amsmath}
\usepackage{cite,color,url}
\usepackage[colorlinks=true
,urlcolor=blue
,anchorcolor=blue
,citecolor=magenta
,filecolor=blue
,linkcolor=blue
,menucolor=blue
,pagecolor=blue
,linktocpage=magenta
,pdfproducer=medialab
]{hyperref}
\def\issue(#1,#2,#3){{\bf #1}, #2 (#3)}

\catcode`\@=11
\def\lsim{\mathrel{\mathpalette\@versim<}}
\def\gsim{\mathrel{\mathpalette\@versim>}}
\def\@versim#1#2{\vcenter{\offinterlineskip
\ialign{$\m@th#1\hfil##\hfil$\crcr#2\crcr\sim\crcr } }}
\catcode`\@=12
\def\baselinestretch{1.20}
\catcode`@=12
\begin{document}

\thispagestyle{empty}
\begin{flushright}
\end{flushright}
\vspace{-0.65in}
\begin{flushright}
SHEP-12-35
\end{flushright}
\begin{center}


{\Large \bf The 125 GeV Higgs signal at the LHC in the 
CP-violating MSSM }\\ 


\vspace{.3in}
{\bf Amit Chakraborty$^a$, Biswaranjan Das$^{b}$,
J. Lorenzo Diaz-Cruz$^c$, Dilip Kumar Ghosh$^{a}$,
 Stefano Moretti$^{d}$ and  Poulose Poulose$^{b}$ } \\[0.25cm]
{\sl $^a$ Department of Theoretical Physics, 
Indian Association for the Cultivation of Science, \\ [-1mm] 
2A \& 2B, Raja S.C.\,Mullick Road, 
Jadavpur, Kolkata 700\,032, India.} \\[0.25cm]
{\sl $^b$ Department of Physics, 
 \\ [-1mm] IIT Guwahati, Assam 781039, India. } \\[0.25cm]
{\sl $^c$ Facultad de Ciencias F\'{\i}sico-Matem\'aticas, \\
Benem\'erita Universidad Aut\'onoma de Puebla, Puebla, 
M\'exico }\\[0.25cm]
{\sl $^d$ School of Physics \& Astronomy, \\
University of Southampton, Highfield, Southampton 
SO17 1BJ, UK.}
\end{center}
\vspace{0.3in}


\begin{abstract}

The ATLAS and CMS collaborations have observed independently 
at the Large Hadron Collider (LHC) a new Higgs-like particle 
with a mass $M_h \sim$ 125 GeV and properties similar 
to that predicted by the Standard Model (SM). 
Although the measurements indicate that this Higgs-like boson 
is compatible with the SM hypothesis, however due to large 
uncertainties in some of the Higgs detection channels, one 
still has the possibility of testing this object as being a 
candidate for some Beyond the SM (BSM) physics scenarios,
for example, the Minimal Supersymmetric 
Standard Model (MSSM), in the CP-conserving version (CPC-MSSM).
In this paper, we evaluate the modifications of these 
CPC-MSSM results when CP-violating (CPV) phases are 
turned on explicitly, leading to the CP-violating MSSM 
(CPV-MSSM). We investigate the role of the CPV phases 
in (some of) the soft Supersymmetry (SUSY) terms on both the 
mass of the lightest Higgs boson $h_1$, and the rates for the 
processes $gg \rightarrow h_1 \rightarrow \gamma \gamma$,
$gg \rightarrow h_1 \rightarrow ZZ^*\rightarrow 4l$,
$gg \rightarrow h_1 \rightarrow WW^*\rightarrow l \nu l \nu$,
$pp \rightarrow V h_1 \rightarrow V b\bar b$ and 
$pp \rightarrow V h_1 \rightarrow V \tau^+\tau^-$,
($V \equiv W^\pm, Z$) at the LHC, considering the impact of the flavor 
constraints as well as the constraints coming from the 
Electric Dipole Moment (EDM) measurements. We find that 
it is possible to have a Higgs mass of about 
125 GeV with relatively small $\tan\beta$, large $A_t$ and a light stop, 
which is consistent with the current SUSY particle 
searches at the LHC. We obtain that the imaginary part of 
the top and bottom Yukawa couplings can take very small 
but non-zero values even after satisfying the recent 
updates from both the ATLAS and CMS collaborations within 
1-2$\sigma$ uncertainties which might be an interesting 
signature to look for at the future run of the LHC. Our 
study shows that the CPV-MSSM provides an equally potential 
solution (like its CP-conserving (CPC) counterpart) 
to the recent LHC Higgs data, in fact offering very little 
in the way of distinction between these two SUSY models 
(CPC-MSSM and CPV-MSSM) at the 7 and 8 TeV run of the LHC. 
Improvement in different Higgs coupling measurements is 
necessary in order to test the possibility of probing the 
small dependence on these CPV phases in the Higgs sector
of the MSSM. 

\end{abstract}


\vskip 1mm
\noindent {\footnotesize E-mail:
{\tt 
\href{mailto:tpac@iacs.res.in}{tpac@iacs.res.in},
\href{mailto:biswaranjan@iitg.ernet.in}{biswaranjan@iitg.ernet.in},
\href{mailto:jldiaz@fcfm.buap.mx}{jldiaz@fcfm.buap.mx},
\href{mailto:tpdkg@iacs.res.in}{tpdkg@iacs.res.in},\\
\href{mailto:S.Moretti@soton.ac.uk}{S.Moretti@soton.ac.uk},
\href{mailto:poulose@iitg.ernet.in}{poulose@iitg.ernet.in}
}
}
\newpage


\section{Introduction}
\label{intro}

The experimental observation of the SM Higgs boson and the 
determination of its properties were among the main 
motivations behind the construction of the LHC. Both the 
ATLAS and CMS collaborations reported a Higgs boson 
discovery with the new particle having mass around 
125 GeV \cite{:2012gu,:2012gk,cms-higgs125,atlas-higgs125}. 
It is also evident
from the results that the observed signals in the 
different production and decay channels available seem to 
follow the SM predictions. However, primarily due to the 
presence of large experimental uncertainties, there could 
be some deviations in some of the individual channels from
the SM expectations.
According to the recent updates on LHC results,
CMS results are consistent with the SM expectations within 
$1\sigma$ uncertainty in all channels, except
for a slight tension in the case of 
$h\rightarrow WW^*$ \cite{cmsnewgam,cmsnewzz,cmsnewww,
cmsnewtautau,cmsnewbb,cmsnewcomb}.
On the other hand, a slight excess still persists in the case 
of ATLAS observations in most of the channels
\cite{atlasnewgam,atlasnewzz,atlasnewww,atlasnewbb,
atlasnewtautau,atlasnewcomb2013,atlasnewcomb}. 
While not incompatible with statistical 
fluctuations,  it is also possible that such deviations could 
signal the presence of BSM physics. For instance, 
one can explain these results in models with an extended Higgs 
sector like those embedded in SUSY 
\cite{susy,Martin:1997ns,Djouadi:2005gi,Djouadi:2005gj}.

SUSY is in fact one of the most popular extensions of the SM, 
with motivations that include: i) the solution to the 
hierarchy and naturalness problems of the SM; ii) the 
unification of the SM gauge couplings at some high scale 
close to the Planck mass; iii) the provision of a Dark 
Matter (DM) candidate (so long that $R$-parity conservation 
is postulated); iv) being a natural ingredient of String 
theories. 

The MSSM though, the simplest realization of SUSY, 
predicts the maximum tree-level value of the lightest 
Higgs mass to be $M_h \leq M_Z$. Significant radiative 
corrections 
are needed in order to push $M_h$ {beyond} the latest 
LEP bound, $M_h > 114$ GeV. However, making the Higgs 
mass  close to  125 GeV requires the inclusion of sizable 
top/stop loop corrections, which depend quadratically on 
the top quark mass and logarithmically on the stop masses, 
combined with a large value of $\tan\beta$, the ratio of 
the Vacuum Expectation Values (VEVs) of the two Higgs 
doublets pertaining to the MSSM. Several studies have 
already been performed in the context of different 
SUSY models, including the MSSM 
\cite{Arbey:2012bp,Bechtle:2012jw,SchmidtHoberg:2012ip,
Heng:2012at,Drees:2012fb,Arbey:2012dq,SchmidtHoberg:2012yy,
Carena:2012xa,Carena:2011aa,Hall:2011aa,Heinemeyer:2011aa,
Arbey:2011ab,Draper:2011aa,Chen:2012wz,Guo:2011ab,He:2011gc,
Djouadi:2011aa,Cheung:2011nv,Hemeda:2013hha,Belyaev:2013rza,
Batell:2011pz} 
(also the constrained version \cite{cmssm125}), 
NMSSM \cite{nmssm125} and  (B--L)SSM \cite{blssm125}. 
All of these scenarios predict a SM-like Higgs boson 
with mass around 125 GeV and also offer solutions explaining 
the potential slight disagreement between the data and the 
SM predictions in different decay channels.  

Another route to follow in order to obtain similar 
results is to consider the possibility of having 
non-zero values of the CPV phases in (some of) the 
soft SUSY parameters that can substantially modify 
Higgs boson phenomenology at colliders at both mass 
spectrum and production/decay level. This motivated 
an avalanche of phenomenological studies in this 
CPV-MSSM framework
\cite{Dedes:1999zh,Dedes:1999sj,Hesselbach:2011nw,
Hesselbach:2009st,Moretti:2007th,Hesselbach:2007en,
Hesselbach:2009gw,Kittel:2004rp,Hinchliffe:2000cn,
Ham:2007gw,Gajdosik:2004ed,Feng:2006ze,Ellis:2004fs,
Baek:1998yn,Altmannshofer:2008hc,Demir:1999ky,Demir:1999hj,
Choi:1999aj,Choi:2001iu,Kane:2000aq,Carena:2000ks,
Ellis:2005ika,Godbole:2006eb,Akeroyd:2001kt,Choi:2001pg,
Choi:2002zp,Deppisch:2009nj,Ghosh:2004cc,Ghosh:2004wr,
Arhrib:2001pg,Bhattacherjee:2012bu}. 
In the presence of CPV complex parameters, 
the top and bottom squarks couplings to the Higgs state  
will be modified substantially in a large domain of the 
MSSM parameter space\cite{Bartl:2003pd,Bartl:2004ws,
Chen:2003bt,Bartl:2003he,Bartl:2003yp}. Conversely 
though the CP phases in the MSSM are significantly 
constrained by the EDM measurements. At the same time, 
the non-zero phases, satisfying the EDM constraints, may 
be allowed, as explained in the following sections, 
and in some details in Refs.  
\cite{Ibrahim:2001ym,Ibrahim:2001ht,Ibrahim:1999af,
Li:2010ax,Romanino:1996cn,Mercolli:2009ns,YaserAyazi:2006zw}.

The Higgs potential of the MSSM is CP invariant at tree level. 
Several studies have been performed to break the CP invariance
of the Higgs potential spontaneously \cite{sponcpv}. However, 
these possibilities are now almost ruled out by various 
experiments \cite{sponcpvexp}. Instead, CP violation can be 
induced explicitly in the Higgs sector of the MSSM. This can 
be achieved by introducing complex parameters that break CP 
invariance in the sfermion and chargino/neutralino sectors.  
There are many new parameters which could in principle be 
complex and thus possess CPV phases, like the Higgsino mass 
parameter ($\mu$), the soft SUSY breaking gaugino masses 
($M_1, M_2, M_3$) and the soft trilinear couplings ($A_f$) 
of the Higgs boson to the (massive) sfermions of flavor $f$. 
In general, each of these phases can be independent. The CPV 
effects are then carried into the Higgs sector through 
the interactions of the two Higgs doublets with the 
sfermions and/or charginos/neutralinos.

In this paper, we will study the possibilities to have the Higgs 
signals with mass around 125 GeV in the context of such a 
CPV-MSSM, which are in agreement with the aforementioned 
LHC data as well as other experimental constraints. 
We will 
look for parameter configurations of the model for which
there exists agreement with both the Higgs mass and the 
rates into the channels observed by the LHC.
We will investigate the dependence of the feasible 
CPV-MSSM signals on the couplings of the Higgs boson to 
both the relevant particle and sparticle states entering 
the model spectrum, as well as upon the masses of the latter, 
thereby aiming at a general understanding of the role of the 
complex phases. While scanning the CPV-MSSM parameter space, 
we also take into account the constraints coming from the 
flavor sector and the EDM measurements.

The paper is structured as follows. In the next section 
we give a brief introduction to the Higgs sector of the 
CPV-MSSM. In Sec.~\ref{constraints} we discuss the relevant 
experimental constraints coming from the SUSY particle searches, 
flavor sector and EDM measurements. In 
Sec.~\ref{parameterspace} we investigate the 
possible numerical values of its parameters after 
performing scans of the CPV-MSSM parameter space against 
available experimental constraints. In Sec.~\ref{lhcresults} 
we present our results on Higgs production and decay 
processes in connection with the LHC Higgs data. Finally, 
we conclude in Sec.~\ref{conclusions}.


\section{A light Higgs mass within the CPV-MSSM }
\label{Thelighthiggs}

Within the framework of the MSSM, non-zero phases 
of $\mu$, $M_i$ ($i=1,2,3$) and/or $A_f$ ($f=t,b,\tau$) 
can induce  
CP violation in the Higgs sector radiatively, via the 
interactions of the Higgs bosons with the sfermions and 
gauginos. These interactions lead to modifications of the 
Higgs masses as well as the Higgs couplings, breaking the 
CP invariance of the tree level scalar potential. Presence 
of CP violation in the Higgs sector leads to 
scalar-pseudoscalar mixing, resulting in CP-mixed physical 
Higgs states. In the following we describe this mixing 
schematically and explicitly present the dependence of 
mixing on different complex parameters. The gauge 
eigenstates of the MSSM Higgs doublets are given by
\begin{equation}
\Phi_1 = 
\left( \begin{array}{c} \phi_1^+ \\ \phi_1^0 + i \eta_1^0 \end{array} \right),
\qquad\qquad
\Phi_2 = 
\left( \begin{array}{c} \phi_2^0 + i \eta_2^0 \\ \phi_2^- \end{array} \right),
\end{equation}
with 
\begin{equation}
<\Phi_1^0>=\frac{1}{\sqrt{2}}\left( \begin{array}{ccc}
0 \\
\upsilon_u
\end{array} \right),\qquad 
<\Phi_2^0>= \frac{1}{\sqrt{2}}\left( \begin{array}{ccc}
\upsilon_d \\
0
\end{array} \right)\nonumber
\end{equation}
and  where $\eta_i^0$ ($i=1,2$) are the pseudoscalar 
components of the two Higgs doublets.

In the presence of the CPV phases in the scalar potential,
the mass matrix for the neutral Higgs bosons enters 
through the general form
\begin{equation}
{\cal L}_{\rm mass} = 
\left( \begin{array}{cc|cc} \eta_1^0 & \eta_2^0 & 
\phi_1^0 & \phi_2^0 \end{array} \right) 
\left( \begin{array}{c|c} {\cal M}_P^2  &  {\cal M}_{SP}^2 \\  
\\  \hline & \\ 
\left[{\cal M}_{SP}^2\right]^T & {\cal M}_S^2 \end{array} 
\right) \left( \begin{array}{c} \eta_1^0 \\ \eta_2^0 \\ \hline \phi_1^0 
\\ \phi_2^0 \end{array} \right). 
\end{equation} 
This $4 \times 4$ mass matrix is divided into $2 \times 2$ 
blocks with ${\cal M}_P^2$ and ${\cal M}_S^2$ representing 
the mixing within the pseudoscalar and scalar states, 
respectively, and the off-diagonal block, ${\cal M}_{SP}^2$, 
representing the mixing between the scalar-pseudoscalar states.
Note that ${\cal M}_{SP}^2$ is absent in the CPC-MSSM and 
is generated in the CPV-MSSM through one-loop 
corrections \cite{Pilaftsis:1999qt,Carena:1995bx,
Pilaftsis:1998pe,Choi:2000wz,Choi:2004kq,Carena:2001fw,
Carena:2000yi,Williams:2011bu,Frank:2006yh}. Different 
contributions to the terms in the 
$2 \times 2$ matrix ${\cal M}_{SP}^2$ 
can be summarized as follows 
\cite{Pilaftsis:1999qt,Carena:1995bx}:
\begin{equation}
{\cal M}^2_{\rm SP} \approx {\cal O}\left ( \frac{M^4_t 
\mid \mu \mid 
\mid A_t \mid}{v^2 32 \pi^2 M^2_{\rm SUSY}}\right ) 
\sin \Phi_{\rm CP} 
\times \left [6, \frac{\mid A_t \mid^2 }{M^2_{\rm SUSY}}, 
\frac{\mid 
\mu\mid^2}{\tan\beta M^2_{\rm SUSY}}, 
\frac{\sin 2\Phi_{\rm CP}\mid 
A_t\mid \mid\mu\mid }{\sin \Phi_{\rm CP} M^2_{\rm SUSY}}
\right ],
\end{equation}
where $\Phi_{\rm CP} = {\rm Arg}(A_t\mu)$, $v = 246$ GeV 
and the mass scale $M_{\rm SUSY}$ is defined by
\begin{equation}
M_{\rm SUSY}^2 = \frac{m^2_{\tilde t_1} + m^2_{\tilde t_2}}{2} 
\ ,
\end{equation}
with ${m_{\tilde t_1}}$ and ${m_{\tilde t_2}}$ being the 
stop masses.

One can easily estimate the degree of CP violation in the 
Higgs sector by considering the dominant one(s) of these 
contributions. For example, sizeable scalar-pseudoscalar 
mixing is possible for a large CPV phase $\Phi_{\rm CP}$, 
$|\mu|$ and $|A_t| > M_{\rm SUSY}$. Apart from a massless 
Goldstone boson $G^0$, which does not mix further with
the other neutral states, the $4 \times 4$ mass matrix 
effectively reduces to a $3 \times 3$ Higgs mass-squared 
matrix  ${\cal M}^2$, in the basis $(A, \phi_1^0, \phi_2^0)$, 
where $A$ is the appropriate eigenstate of ${\cal M}_P^2$.  
The $3\times 3$ symmetric matrix ${\cal M}^2_{ij}$ can be 
diagonalized by an orthogonal matrix ${\cal O}$, i.e., 
$M^2_i \delta_{ij} = O_{ik} {\cal M}^2_{kl} O_{jl}$,
leading to physical states, $h_i= O_{ji}~\phi_j$, where 
$\phi_j \equiv (A, \phi_1^0, \phi_2^0)$. In this article, 
the physical mass eigenstates $h_1, h_2 $ and $h_3$ are 
considered in ascending order of mass 
($M_{h_1}<M_{h_2}<M_{h_3}$).
Moreover, as $A$ is no longer a physical state, the charged 
Higgs boson mass $M_{H^\pm}$ is a more appropriate parameter 
for the description of the CPV-MSSM Higgs sector instead of 
$M_{A}$ often used in the CPC-MSSM. Hence, the tree level 
Higgs masses in the CPV-MSSM can be conveniently expressed 
in terms of $\tan\beta$ and $M_{H^\pm}$. 

Radiative corrections enhance the Higgs mass significantly 
via the top quark Yukawa coupling, the third generation top 
squark mass parameters $M_{Q3}, \ M_{U3}$ and the trilinear 
coupling $A_t$, while the bottom squark sector has a somewhat 
subdued effect. At the same time though, flavor physics 
observations from the $b$-quark sector often serve as 
stringent constraints on Higgs phenomenology and we therefore 
include the sbottom sector parameters $M_{D3}$ and the 
trilinear coupling $A_b$ along with the above mentioned top 
squark parameters. The stau sector, in principle, can play a 
significant role in the Higgs to di-photon decay mode. 
To take this into account, we include the parameters 
$M_{L3}, \ M_{E3}$ and $A_\tau$ corresponding to the stau 
sector in our parameter space scan.

Coming to the first two generations of the 
soft masses, it is well known that they have very little effect 
on the Higgs sector of the MSSM. At the same time, their phases 
$\phi_{A_{e/\mu}}$, $\phi_{A_{u/d}}$ can provide significant 
contributions to the atomic EDMs. These can however be 
drastically reduced either by assuming these phases to be
sufficiently small or by taking the first and second generation 
squarks and sleptons sufficiently heavy. This is achieved by 
setting the hierarchy factor between the first two and  
third generation soft masses to be 20. Nonetheless, sizeable 
contributions to the EDMs are always 
possible from Higgs-mediated two-loop diagrams
\cite{Chang:1998uc,Pilaftsis:2002fe,Olive:2005ru}. Therefore, 
in order to ascertain whether the regions of parameter space 
of interest here are potentially compatible with the EDM 
constraints, we have calculated the EDMs of Thallium, 
Mercury, electron and 
neutron ($d_{Tl}, d_{Hg}, d_e, d_n$) and compared 
the results with the current bounds 
\cite{Li:2010ax,Griffith:2009zz,Hudson:2011zz,Regan:2002ta,
Jung:2013mg,Baron:2013eja,Jung:2013hka,Serebrov:2013tba,
Baker:2006ts,Ellis:2011hp,Schiff:1963zz,Engel:2013lsa}. However, 
we would like to clarify that  in the present analysis, wherein 
we mainly focus on the possibility of a 125 GeV Higgs boson 
signal in the CPV-MSSM, we do not perform detailed
studies of the different EDMs. In fact, in general, it has been
shown in \cite{Li:2010ax} that the constraints from the EDMs are
highly dependent upon the combinations of different phases
of soft SUSY breaking parameters as different loop diagrams 
can interfere either destructively or constructively so as 
to either suppress or enhance, respectively, individual 
contributions to the EDMs. In the case of $d_{Hg}$, the 
experimental limits put severe constraint on $\phi_3$, due 
to the strong correlation between this phase and 
{$\phi_{A_{u,d}}$}, both of which enter the EDM operators 
at one loop level. In contrast, $d_n$ and $d_{Tl}$ limits have 
a relatively stronger impact on $\phi_2$ though, presently, 
the latter constraint is not strongly correlated with any of 
the other phases. A detailed analysis of the impact of the 
EDM data in the Higgs sector of the CPV-MSSM, considering 
all the three Higgs bosons 
and all CPV phases, is complicated and beyond 
the scope of this study. Nevertheless, herein we take 
a straightforward approach and present our results after 
comparing with the available EDM constraints. In the following 
section we present our numerical analyses and discuss 
the parameter space within the CPV-MSSM respecting these 
constraints along with all other experimental restrictions 
including those from the flavor sector.


\section{Current experimental constraints}
\label{constraints}

As explained in Sec.~\ref{Thelighthiggs}, the non-trivial 
CPV  phases modify the Higgs mass significantly by 
introducing mixing between the scalar and pseudoscalar 
Higgs sector. CPV phases can also affect the Higgs couplings 
with the gauge bosons and fermions, altering significantly 
their tree level values. For example, in a situation with 
maximal CP violation, known as CPX scenario 
\cite{Carena:2000ks,Carena:2000yi,Sopczak:2006vn,Schael:2006cr}, 
one can have the lightest Higgs boson which is almost 
CP-odd with a highly suppressed coupling to a pair of 
$W$'s or $Z$'s \cite{Pilaftsis:1999qt,Carena:1995bx}. 

Both the ATLAS and CMS collaborations reported the best 
fit results of invariant mass in the two high
resolution channels, $\gamma\gamma$ and 
$Z{Z^*} \to 4\ell$ ($\ell = e,\mu$) as 
125.5 $\pm$ 0.2 (stat.) $^{+0.5}_{-0.6}$ (syst.) 
GeV ~{\cite{atlas-higgs125}} and 125.3 $\pm$ 0.4 (stat.) 
$\pm$  0.5 (syst.) GeV {\cite{cms-higgs125}}, respectively.
Thus, considering the 1$\sigma$ uncertainty band around 
the best fit value, we primarily demand that the lightest 
Higgs boson mass $(M_{h_1})$ should always lie in the range
of 124.0 - 126.0 GeV, while scanning the CPV-MSSM parameter 
space. Besides this, we also enforce the following 
constraints to select the final allowed parameter space 
points for our further analyses.

We impose 95\% Confidence Limit (CL)\footnote{In our analysis, 
we consider 3$\sigma$ bound for almost 
all the experimental constraints. But, for the sparticle masses,
we find 95\% CL limit from the Particle Data Group~\cite{PDG}. 
We check that the updated results on different 
sparticle masses, which are available in the literature but not yet included in the 
PDG database, do not change our results substantially.}
lower bounds on the masses of sparticles, listed by the 
Particle Data Group (PDG)~\cite{PDG}, as follows: 
\begin{eqnarray}\label{eq:C}
M_{\widetilde{\chi}_1^0} > 46 ~{\rm{GeV}}, \,~~ M_{\widetilde{\chi}_2^0} > 62.4~{\rm{GeV}}, \,~~ M_{\widetilde{\chi}_1^\pm} > 94  ~{\rm{GeV}}, \nonumber \\
M_{\tilde{t}_1} > 95.7  ~{\rm{GeV}}, \,~~ M_{\widetilde{b}_1} > 89  ~{\rm{GeV}}, \,~~ M_{\widetilde{g}} > 800  ~{\rm{GeV}}. 
\end{eqnarray}

It is well known that the flavor observables play 
a crucial role in determining the viable regions of the 
SUSY parameter space. Several rare b-decays, which are 
helicity suppressed in the SM, can acquire substantial 
contribution from different SUSY particles present 
in the model and these corrections may come with 
same or opposite sign with the SM expectations. To take 
into account the stringent constraints on the SUSY 
parameter space coming from the flavor sector, we 
consider several low energy processes like the purely leptonic decay of  
$B_s \rightarrow \mu^+ \mu^-$ and 
$B_d \rightarrow \tau^+ \tau^-$, the 
radiative decay $b \rightarrow s \gamma$ etc. The 
$B_s \rightarrow \mu^+ \mu^-$ decay is a flavor changing 
neutral current process which occurs at the loop level in 
both the SM and MSSM. In the SM, it is helicity suppressed 
by the muon mass, which results in tiny SM expectation for 
the branching ratio of the order of 
$10^{-9}$ \cite{Buras:2012ru}. For large values of 
$\tan\beta$, order of magnitude enhancements of the 
$Br(B_s \rightarrow \mu^+ \mu^-)$ are possible in the MSSM, 
for details see Ref.\cite{Babu:1999hn,
Choudhury:1998ze,Altmannshofer:2012ks} 
and the references therein. In the MSSM, the dominant 
contribution mainly comes from the Higgs penguin diagrams 
with the exchange of the heavy scalars present in the flavor 
changing b$\to$s couplings. Besides, there are also 
contributions from the charged Higgs and gluino exchange 
diagrams which may interfere constructively or destructively 
with the Higgs diagrams and the SM expectations depending 
upon the sign of the $\mu$ and $A_t$ terms. Due to its 
strong dependence on $\tan\beta$, MSSM parameter 
space with large $\tan\beta$ is now highly constrained 
by the current experimental results on 
$Br(B_s \to \mu^+ \mu^-)$
\cite{TheATLAScollaboration:2013wia,CMS:2013hja}.
The combined experimental result from the LHCb and CMS for 
$Br(B_s \to \mu^+ \mu^-)$ is $(2.9\pm­0.7) \times 10^{-9}$
~\cite{CMSandLHCbCollaborations:2013pla,Aaij:2013aka,
Chatrchyan:2013bka} and in our analysis we consider 
the 3$\sigma$ error band around the central value, 
\begin{equation}
0.8 \times 10^{-9} < Br(B_s \to \mu^+ \mu^-) < 5.0 \times 10^{-9}.
\label{bsmumulimit}
\end{equation}
Let us now consider another important b-observable, 
namely $Br(b \rightarrow s \gamma)$. In the SM, it comes 
from the $t-W$ loop\cite{bsgammaSMoriginals} and, in the 
MSSM, the dominant contribution comes from the 
$t-H^\pm$ and ${\tilde t}_{1,2}-{\tilde \chi}_{1,2}^\pm$ 
loops\cite{bsgammaSUSYorigEtc}, where the former have the 
same sign with the SM $t-W$ loop. The chargino loop 
contribution is proportional to the product 
$A_t \mu \tan\beta$.  Depending on the sign of $A_t \mu$, 
there might be cancellation or enhancement between the above 
two loop contributions within the MSSM\cite{Carena:2000uj}. 
Here, we choose positive $A_t$ and positive $\mu$ and so we 
expect some cancellation between these different SUSY 
corrections for large values of $\tan\beta$. Considering the 
large uncertainty in the measurement of 
$b \rightarrow s \gamma$, we here assume 3$\sigma$ 
uncertainty around the experimental value of 
$Br(b \rightarrow s \gamma)=(3.43 \pm 0.22)
\times 10^{-4}$\cite{Amhis:2012bh} which leads to
\begin{equation}
2.77 \times 10^{-4}  <Br(b \rightarrow s  \gamma)<4.09 \times 10^{-4}.
\label{bsgammalimits}
\end{equation}

Apart from the above-mentioned two flavor constraints, 
which play a significant role in the present study, there 
exist other flavor constraints with subdued influences. 
The mass differences measured in the $B_{0} - \bar{B_0}$ 
mixing, $\Delta M_{B_d}$ and $\Delta M_{B_s}$, are 
equally sensitive to the new physics contributions. For both these two 
observables, at large $\tan\beta$, non-negligible 
contribution comes from the most dominant double scalar 
penguin (DP) diagrams \cite{Buras:2002vd,Ball:2006xx}. 
The experimental and the SM values for the mass differences 
in the $B_d$ system are: 
$\Delta M_{B_d}^{\rm Exp}$ = $0.510 \pm 0.004$ ${{\rm ps}^{-1}}$~\cite{PDG,Isidori:2014rba,Amhis:2012bh} 
and $\Delta M_{B_d}^{\rm SM}$ = $0.502 \pm 0.006$ ${{\rm ps}^{-1}}$~\cite{Bona:2006ah}, respectively. 
On the other hand, for the $B_s$ system they are, $\Delta M_{B_s}^{\rm Exp}$ = $17.768 \pm 0.024$ ${{\rm ps}^{-1}}$
~\cite{Isidori:2014rba,Aaij:2013mpa} 
and $\Delta M_{B_s}^{\rm SM}$ = $17.3 \pm 2.6$ ${{\rm ps}^{-1}}$~\cite{Lenz:2011ti}. The SUSY contributions to the 
$B_d^0-\bar B_d^0$ and $B_s^0-\bar B_s^0$ 
mass differences are usually denoted as 
$\Delta M_{B_d}^{\rm SUSY}$ and 
$\Delta M_{B_s}^{\rm SUSY}$, respectively and calculated by 
subtracting the SM prediction from the experimentally 
measured quantity i.e. 
$\Delta M_{B_d}^{\rm SUSY} = \Delta M_{B_d}^{\rm Exp}- \Delta M_{B_d}^{\rm SM}$ 
\cite{Isidori:2006pk,Fornengo:2010mk}. Note that, the 
theoretical uncertainty associated to $\Delta M_{B_s}$ 
dominates the experimental uncertainties, unlike 
$\Delta M_{B_d}$ where both theoretical and experimental 
error bars are relatively small. In rest of 
our analysis, we therefore, consider only $\Delta M_{B_d}$ mass 
difference\footnote{We check that if we consider the 
$\Delta M_{B_s}$ mass difference after 
satisfying eq.\ref{eq:D} and allow even 1$\sigma$ uncertainty 
in $\Delta M_{B_s}^{\rm SUSY}$ estimation, we loose 
only $\sim$1\% points. We plot the correlation of these 
two observables ($\Delta M_{B_s}^{\rm SUSY}$ and 
$\Delta M_{B_d}^{\rm SUSY}$) in Fig.\ref{fig:7}(a) 
and the figure itself justify our claim.} and allow the SUSY contribution 
$\Delta M_{B_d}^{\rm SUSY}$ lie within the 3$\sigma$ 
error band drawn around the experimental best-fit number.   
We also consider the ratio of the experimentally 
measured $Br(B_u \to \tau \nu)$ to its SM value, 
$R_{B_u\to\tau \nu}$ = $\frac{Br^{\rm Exp}(B_u \to \tau \nu)}{Br^{\rm SM}(B_u \to \tau \nu)}$ = 
$1.21 \pm 0.30$~\cite{Altmannshofer:2012ks,Chakraborti:2012up}.
Besides, we further check whether our results are consistent 
with the experimental result on the 
$Br(B_d \to \tau^{+} \tau^{-})$ available in the 
Particle Data Group \cite{PDG}. 

Finally, we consider the measurement of direct CP 
asymmetry, $A_{\rm CP}(B \to X_s \gamma)$ associated with 
the $B \to X_s \gamma$ decay with its present limit 
$-0.008 \pm 0.029$~\cite{PDG}. So, in summary, the 
experimental limits used, corresponding to $3\sigma$ 
uncertainty, (except the $Br(B_d \to \tau^{+} \tau^{-})$ 
which is given at 90\% CL in the PDG) are the following:
\begin{eqnarray}\label{eq:D}
 0.31 < R_{B_u\to\tau \nu} < 2.1, \,~~  Br(B_d \to \tau^{+} \tau^{-}) < 4.1 \times 10^{-3},\nonumber \\ 
 -0.095 < A_{\rm CP}(B \to X_{s} \gamma) < 0.079, \,~~  -0.0136 < \Delta M_{B_d}^{\rm SUSY} < 0.0296 ~{\rm{ps^{-1}}}. 
\end{eqnarray}
All these constraints are imposed on the 
points satisfying the primary selection criterion on the 
Higgs boson mass.  

Furthermore, we investigate the effect of the EDM constraints on 
the parameter space. 
As already discussed, atomic EDMs receive contributions from 
$d_e$, as well as quark EDMs or Chromo EDMs (CEDMs). 
While in a quantum field 
theory, like the CPV-MSSM being discussed here, the presence of 
CPV phases induces EDMs for elementary particles like the 
electrons and quarks. How these contributions present 
themselves at the atomic level is a complex phenomenon, which 
depends on the nature of the atom or molecule being studied 
and on the theoretical model being considered. In the case of 
diamagnetic systems like Mercury (Hg), the dominant contribution 
comes from the CEDMs, while the effect of $d_e$ is sub-dominant.
It is known that these atomic EDMs receive large theoretical 
(hadronic and nuclear) uncertainties arising from the hadronic 
CPV. Besides, while the EDM of Hg is one of the best known 
experimentally, theoretical calculations using different 
techniques do not quite agree with each 
other, for reasons those are not fully understood 
\cite{Engel:2013lsa}.
Thus, the upper bounds on $d_e$ obtained from these results 
should be considered with caution \cite{Jung:2013mg,
Jung:2013hka}. In contrast, in the case 
of paramagnetic systems like Thallium (Tl) and 
Ytterbium Fluoride (YbF), the atomic EDMs depend 
on $d_e$ and another term arising 
from electron-nucleon interactions, therefore the $d_e$ 
extracted from these systems are more reliable. 
Traditionally, while extracting $d_e$ from these systems, 
it is assumed that only the single unpaired electron 
would contribute to their EDMs. Besides, there are also 
direct measurements on $d_n$
\cite{Serebrov:2013tba,Baker:2006ts}, which receives 
contributions from the CEDMs arising in several BSM models. 
Similar to the atomic case, these results too receive large 
theoretical uncertainties. We listed the current bounds 
on $d_n$, $d_{Tl}$ and $d_{Hg}$ in Tab.\ref{tab:edm}.
\begin{table}[h]
\vspace*{5mm}
\centering
\begin{tabular}{|c|c|}
\hline
System & Present limit on absolute value \\
\hline\hline
$\mid d_n\mid$ & 3.3 $\times 10^{-26}$ e cm (95\% CL) \cite{Serebrov:2013tba,Baker:2006ts} \\ [-2mm]
$\mid d_{Tl}\mid$ & 9.0 $\times 10^{-25}$ e cm (90\% CL) \cite{Li:2010ax,Regan:2002ta} \\ [-2mm]
$\mid d_{Hg}\mid$ & 3.1 $\times 10^{-29}$ e cm (95\% CL) \cite{Griffith:2009zz} \\
\hline
\end{tabular}
\def\baselinestretch{1.1}
\caption{Summary table for the current experimental limits on 
$d_n$, $d_{Tl}$ and $d_{Hg}$.}
\def\baselinestretch{1.0}
\label{tab:edm}
\end{table}
At present, the most stringent model independent limit on $d_e$ 
stem from the searches for the EDMs of YbF and Tl, with upper 
limits of $1.05 \times 10^{-27}$ e cm \cite{Hudson:2011zz}
and $1.6 \times 10^{-27}$ e cm \cite{Regan:2002ta} at
90\% CL, respectively. An improved analysis including the 
effect of electron-nucleon interaction and combining the 
results from Tl, YbF and Hg is available in Ref.
\cite{Jung:2013mg}. While considering the $d_e$ constraints 
coming from these experiments, we adopt the result of this 
analysis with $d_e$ given at 95\% CL \cite{Jung:2013mg} as
\begin{eqnarray}\label{eq:Q}
\mid d_e\mid < 1.4 \times 10^{-27}~~{\rm e~cm}.
\end{eqnarray} 
However, the Advanced Cold Molecule Electron (ACME) 
EDM Collaboration \cite{Baron:2013eja} measurement 
recently put a strong limit on $d_e$ which is down by one 
order of magnitude compared to the previous measurements. 
The experimental bound at 90\% CL is:\footnote{Here we would 
like to note that there are certain 
observables for which 90\% or 95\% CL data are  
available in the literature, so we consider
them at the same CL which are available, for example the 
sparticle mass bounds at the PDG are given at 95\% CL, 
while some of the EDM results are available
either at 90\% CL or at 95\% CL.}
\begin{eqnarray}\label{eq:A}
\mid d_e\mid < 8.7 \times 10^{-29}~~{\rm e~cm}.
\end{eqnarray}
In the next section,  we first discuss the details 
of our CPV-MSSM parameter space scan and then show the 
impact of these constraints on the CPV-MSSM parameter space.


\section{Impact of the constraints on the parameter space}
\label{parameterspace}

Here we explore the CPV-MSSM parameter space in order 
to estimate the regions which respect all the above 
mentioned experimental constraints. The scans were 
performed using the publicly available numerical package CPsuperH (version 2.3) 
\cite{cpsuperh}. We here consider two seperate scans 
with different sets of input parameters.\footnote{The reader may note 
that the first set of the parameter space scanning was 
planned and performed when preliminary results of the 
LHC were available. However, the second scan is performed 
keeping in mind the latest results of the LHC and the 
recent electron EDM measurement by the ACME collaboration.} The choice 
of the first set of input parameters was aimed at 
maximizing the effect of CP violation in MSSM, 
while the second set was with a view at searching for 
solutions within the CPV-MSSM compatible with the LHC and 
other experimental results.


\subsection{Scan 1: with maximum CPV phases}
\label{pspace-firstscan}
The first scan considered the following values of the 
input parameters:
\begin{eqnarray}\label{eq:S}
1 < \tan\beta < 60,&\ \  100~ {\rm GeV} < M_{H^\pm} < 300~ {\rm GeV}, \nonumber \\
50~ {\rm GeV} < |M_1| < 500~ {\rm GeV},&\ \  100~ {\rm GeV} < |M_2| < 1000~ {\rm GeV}, \nonumber \\
500~ {\rm GeV} < A_{\rm t} = A_{\rm b} = A_{\rm \tau} < 3000~ {\rm GeV}, & \ \ 
500~ {\rm GeV} <~|\mu| ~~ < 2000~ {\rm GeV}, \nonumber \\
500~ {\rm GeV} < M_{\rm Q3},~M_{\rm U3} < 2000~ {\rm GeV}, & \ \ 500~ {\rm GeV} < M_{\rm D3}~< 2000~ {\rm GeV} , \nonumber \\
100~ {\rm GeV} < M_{\rm L3},~M_{\rm E3} < 2000~ {\rm GeV}. 
\label{parameterRanges}
\end{eqnarray}

The 100 GeV lower limit for $M_2$ is taken from the LEP-2 
lower bound from model-independent chargino searches, while 
the lower limit on $\tan\beta$ is close to the LEP-2 Higgs 
search exclusion. In Ref.~\cite{Hesselbach:2007en,Hesselbach:2009gw}, it was shown 
that there is a transition point at $M_{H^\pm} \sim 150$ GeV (for 
some specific choices of the model parameters) above which 
the lightest Higgs mass state, $h_1$, is almost a pure scalar 
state and thus there would be no CP violation effect through 
the scalar and pseudoscalar mixing. So, in our first scan (Scan 1), we set 
the upper limit on $M_{H^\pm}$ at 300 GeV in accordance with the above observation.
However, we will see in Sec.~\ref{pspace-secondscan}, interesting 
phenomenology appears when we relax this upper limit on the $H^\pm$ mass. 
In order to have maximum CP violation, 
the three phases 
$\phi_{A_f}  (f = t,b,\tau )$ and $\phi_{3}$ are fixed at 
$90^{\rm o}$, while all other phases are set to zero.
Trilinear couplings of the first and second sfermion families 
($|A_e|,|A_{\mu}|,|A_u|,|A_d|,|A_c|,|A_s|$) are less relevant 
for our analysis, hence we have set them to zero. Both the 
CMS and ATLAS collaborations have already excluded gluino 
masses less than 1.1 TeV for different possible final states 
in the context of the Constrained-MSSM (CMSSM) 
\cite{CMS-PAS-SUS-12-005,ATLAS-CONF-2012-109}. Note that, the 
CMSSM bound can not be directly applied here as the bound is 
expected to change in the context of the CPV-MSSM due to 
modifications in the different decay/branching ratios. Here, 
we fix the magnitude and phase of the gluino mass parameter 
$M_3$ at 1.2 TeV and $90^o$ respectively just to reduce the 
number of free parameters.

\begin{figure}[!htb]\centering
\begin{tabular}{c c}
\includegraphics[angle=0,width=70mm]{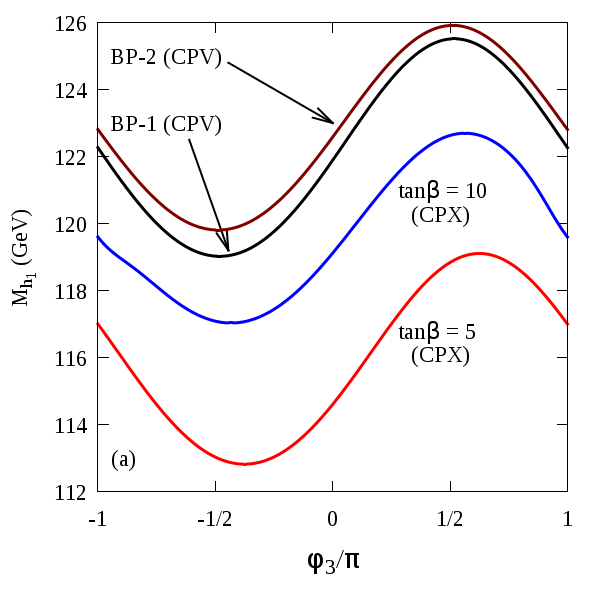} &
\hspace{1.5cm}
\includegraphics[angle=0,width=70mm]{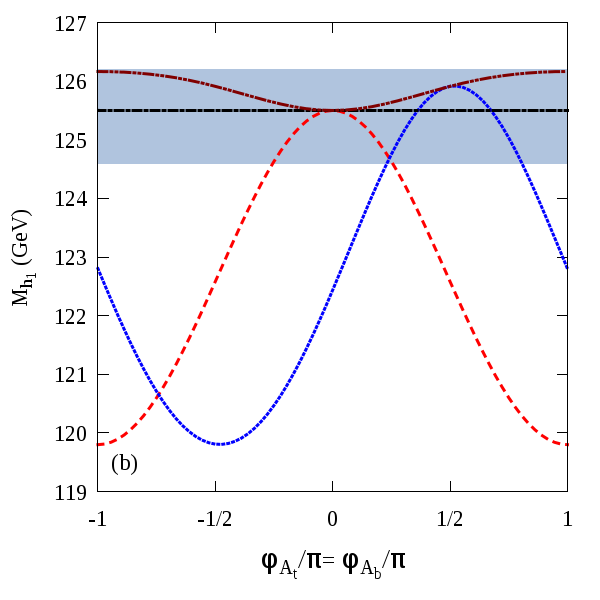} 
\end{tabular}
\caption{In panel (a) we show the dependence of 
$M_{h_1}$ upon $\phi_3$ for two 
BPs from our scan (upper two curves) 
against two benchmark points in the CPX
scenario (lower two curves), keeping $\phi_{A_t} = \phi_{A_b} = \pi/2 $. 
While in panel (b) the red (dash-dotted), 
blue (small dash) and 
brown (long-dash dotted) curves correspond to 
$\phi_3 = 0$, $\phi_3 = \pi/2$, and $\phi_{A_t} = \phi_{A_b}= \phi_3 $, 
respectively, corresponding to BP-2.
The horizontal solid curve in panel (b) represents the value of 
$M_{h_1}$ in the CPC-MSSM and the shaded region corresponds
to the $1\sigma $ range of the observed Higgs boson mass 
125.3 $\pm$ 0.4 (stat.) $\pm$  0.5 (syst.) GeV by the 
CMS collaboration {\cite{cms-higgs125}}.}
\label{fig:1}
\end{figure}

We now begin our discussion on the numerical analyses by 
defining some Benchmark Points (BPs). In Fig.\ref{fig:1}(a),
we display the variation of the lightest Higgs boson 
mass $(M_{h_1})$ as a function of the phase $\phi_{3}$, 
while keeping $\phi_{A_f} = \pi/2$ ($f=t,b,\tau$). 
The upper two curves represent two characteristic 
BPs (BP-1 \& BP-2, see Tab.\ref{tab:BP12}) obtained from 
the scan of the CPV-MSSM parameter space represented 
by eq.(\ref{eq:S}), whereas the lower two curves represent 
the CPX scenario \cite{Carena:2000ks,Carena:2000yi,Sopczak:2006vn,Schael:2006cr}. In Fig.\ref{fig:1}(b), we present a 
similar variation of the lightest Higgs boson mass  
$M_{h_1}$ with the CPV phase but this time with different 
combinations of phases for the BP-2 as displayed in 
Tab.\ref{tab:BP12}. From Fig.\ref{fig:1}, it is clear that 
the mass of the Higgs boson crucially depends on 
the CPV phases. In particular, notice that the radiative 
corrections to the lightest Higgs boson $h_1$ mass 
strongly depend on the stop mixing parameter 
$X_t = A_t - \mu  \cot \beta $. Now, in our case, 
$\mu $ is real while $A_t$ is a complex 
quantity. Hence, for different choice of phases, $X_t$ can 
change, resulting in significant variations of the Higgs mass 
as illustrated by the red (dash-dotted) and blue (small-dash) 
curves. 
\begin{table}[h]
\centering 
\begin{tabular}{|c|cccccccccc|c|}
\hline
BP & $M_1$ & $M_2$ & $M_3$ & $\tan\beta$ & $M_{H^\pm}$ & $M_{Q3}$ & $M_{D3}$ & $M_{L3}$ & $A_t$
& $\mu$ & $M_{h_1}$ \\ 
\hline\hline
1 & 496.7 & 356.1 & 1200 & 7.7 & 268.6 & 758.2 & 889.3 & 125.6 & 2458.0 &  796.6 & 125.5 \\ [-2mm]
2 & 469.7 & 456.8 & 1200 & 9.4 & 294.2 & 1371.8 & 704.5 & 1221.6 & 2621.7 & 1179.2 & 125.9\\
\hline
\end{tabular}
\def\baselinestretch{1.1}
\caption{Two BPs obtained after performing a random scan over 
the CPV-MSSM parameter space using CPsuperH. In addition to 
the parameters relevant to describe the Higgs sector,  
we have also presented the mass of the lightest neutral 
Higgs boson $h_1$. We have fixed the CPV phases to 
90$^{\rm o}$. All masses, $A_t$ and $\mu$ are expressed in
 units of GeV.}
\def\baselinestretch{1.0}
\label{tab:BP12}
\end{table}

In Tab.\ref{tab:BP12}, we have listed the details of BP-1 
and BP-2, where the last column gives the mass of the 
lightest neutral Higgs boson. These two points are 
illustrative of the fact that  it is always possible to 
have the lightest Higgs boson mass around 125 GeV when 
one includes the sizeable corrections coming from the CPV 
phases onto the MSSM results. 

\begin{figure}[!htb]\centering
\vspace*{7mm}
\begin{tabular}{c c}
\includegraphics[angle=0,width=70mm]{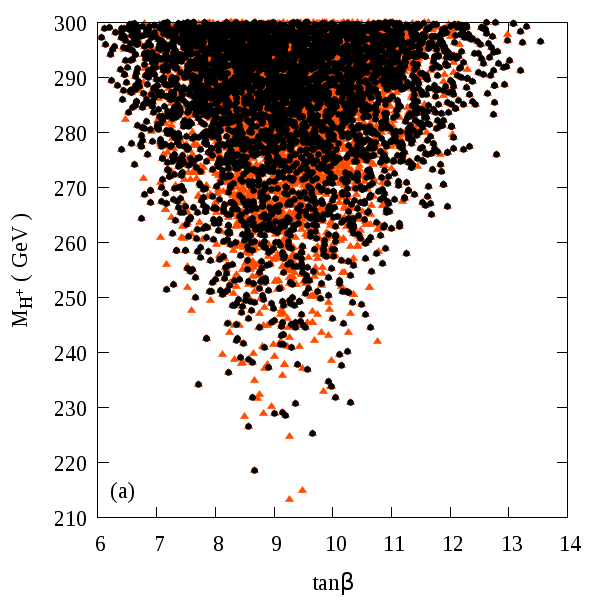} &
\hspace{1.5cm}
\includegraphics[angle=0,width=70mm]{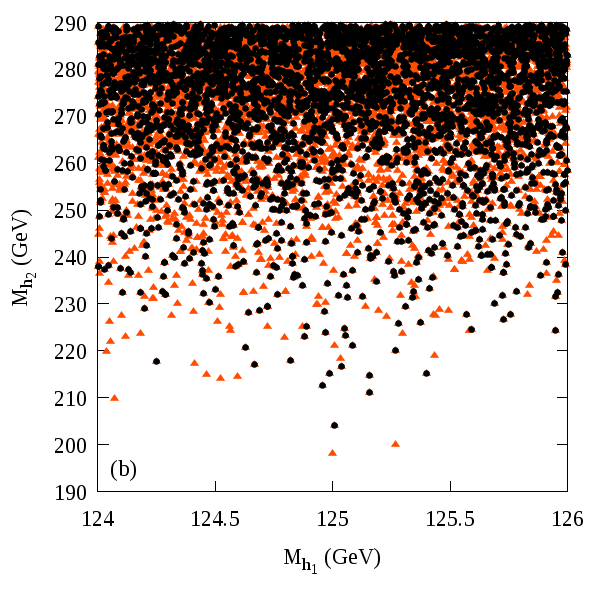} 
\end{tabular}
\vspace*{-1mm}
\caption{Constraints on the (a)
$\tan\beta - M_{H^\pm}$ plane
and (b) $M_{h_1} - M_{h_2}$ plane
obtained after scanning the CPV-MSSM parameter space randomly.
The brick red/grey triangles
are allowed by the set of constraints mentioned in
eq.(\ref{eq:C}) to eq.(\ref{eq:D}) while black 
points represent points which in addition to those constraints
also satisfy the EDM constraint given in eq.(\ref{eq:Q}).} 
\label{fig:2}
\end{figure}

Before proceeding to analyze some LHC observables, it is 
important to take a look at the CPV-MSSM parameters and 
the particle spectrum after constraints are enforced upon 
the points obtained in the random scan with eq.(\ref{eq:S}). 
In Fig.\ref{fig:2}(a) and (b) we have shown the allowed 
region in the 
$\tan\beta - M_{H^\pm} $ and $M_{h_1}-M_{h_2}$ planes, 
respectively. The brick red/grey points are allowed by the 
set of constraints mentioned in
eq.(\ref{eq:C}) to eq.(\ref{eq:D}) while the black dots 
represent points which in addition to those constraints
also satisfy the EDM constraint given in eq.(\ref{eq:Q}).
We find that once we impose the $d_{e}$ constraint as given 
in eq.(\ref{eq:Q}), the allowed parameter space shrinks to 
the region depicted by the black dots, still with sufficient 
regions in the parameter space surviving all the constraints. 
From Fig.\ref{fig:2}(a) one can conclude that the current 
limit on $Br(B_s \to \mu^+ \mu^-)$ prefers low to medium 
values of $\tan\beta \sim 6-13$ and a somewhat heavy  
charged Higgs mass $M_{H^\pm} \gsim 200 $ GeV 
(brick red/grey points). In the CPC-MSSM, one has 
${Br}(B_s \to \mu^+\mu^-) \propto \tan\beta^6/M_A^4 $,
where $M_A$ is the pseudoscalar Higgs boson mass. Hence, to 
satisfy the current limit on ${Br}(B_s \to \mu^+ \mu^-)$, 
one requires a heavier $M_{H^\pm}$ and a lower $\tan\beta $. 
One may expect modifications in this formula in the presence 
of CPV phases, however phases will not change it 
significantly. Fig.\ref{fig:2}(b) shows that, when the 
lightest Higgs boson is SM-like, the second lightest Higgs 
boson can have mass around 200--300 GeV. This mass window 
may be accessible at the 14 TeV LHC run 
via $ g g \to h_2 \to h_1 Z $, followed
by $ h_1 \to b {\bar b}$ and 
$Z \to \nu \bar\nu~{\rm or}~\ell^+ \ell^- $, as 
the analysis is very similar to the one performed in the 
case of the heavy Higgs boson searches at the LHC in the 
CPC-MSSM \cite{Djouadi:2005gj}.

We further proceed to check how the most updated 
$d_e$ measurements affect the scanned CPV-MSSM parameter space. 
We find that, the latest bound on the $d_{e}$, quoted in 
eq.(\ref{eq:A}), completely negates the parameter space 
region corresponding to eq.(\ref{eq:S}). However, we shall 
see that relaxing the CPV phases from their maximal values 
($90^o$), and with more appropriate choices of the parameters, 
one can satisfy the latest $d_e$ bound along with all other 
EDM constraints (mentioned in Tab.\ref{tab:edm}) and the low 
energy experimental bounds. Without moving further with the 
results corresponding to this set of parameters
(eq.(\ref{eq:S})), 
we now proceed for a new scan with modified parameter set, 
in order to satisfy all the constraints discussed before 
including the stronger electron EDM bound. 


\subsection{Scan 2: allowing CPV phases to vary in the 
range of $0^o-90^o$} 
\label{pspace-secondscan}

Moving away from the maximal CPV scenario
(where $\phi_3$ and $\phi_{A_f}$ are fixed to $90^o$), 
sample test scans varying the magnitude of $M_3$ and the 
CPV phases ($\phi_3$, $\phi_{A_t}$, $\phi_{A_b}$ and 
$\phi_{A_{\tau}}$) between $0^o-90^o$, were performed to 
optimize the parameter ranges suitable to accommodate all 
the experimental constraints coming from the flavor sector 
and the EDM measurements. After a dedicated analysis using 
those sample data sets, we fix the ranges (upper and lower 
limits) of the CPV parameters and then proceed to scan the 
CPV parameter space for larger statistics. In our second 
scan, we choose an extended $M_{H^\pm}$ mass range by setting 
the upper extreme to 1000 GeV, which was fixed at 300 GeV 
in our first scan, in order to avoid the loss of significant 
amount of parameter space due to the 300 GeV upper 
bound on $M_{H^\pm}$. Similarly for $\tan\beta$ we now 
choose the range to be from 1 to 30, as our first scan has 
already discarded regions with very large values 
of $\tan\beta$.

We consider the following set of parameters in our final scan 
and vary them randomly within the specified ranges:
\begin{eqnarray}\label{eq:B}
1 < \tan\beta < 30,&\ \ 250~ {\rm GeV} < M_{H^\pm} < 1000~ {\rm GeV}, \nonumber \\
50~ {\rm GeV} < |M_1| < 500~ {\rm GeV},&\ \  100~ {\rm GeV} < |M_2| < 1000~ {\rm GeV}, \nonumber \\
800~ {\rm GeV} < |M_3| < 2000~ {\rm GeV},&\ \ 500~ {\rm GeV} <~|\mu| ~~ < 1000~ {\rm GeV}, \nonumber \\
1500~ {\rm GeV} < A_{\rm t} < 3000~ {\rm GeV},&\ \ 500~ {\rm GeV} < A_{\rm b}, A_{\rm \tau} < 3000~ {\rm GeV}, \nonumber \\ 
500~ {\rm GeV} < M_{\rm Q3} < 1500~ {\rm GeV}, & \ \ 1000~ {\rm GeV} < M_{\rm U3}~< 3000~ {\rm GeV}, \nonumber \\
500~ {\rm GeV} < M_{\rm D3} < 2000~ {\rm GeV}, & \ \ 100~ {\rm GeV} < M_{\rm L3},~M_{\rm E3} < 2000~ {\rm GeV}, \nonumber \\
0^o < \phi_3, \phi_{A_t}, \phi_{A_b}, \phi_{A_\tau} < 90^o. 
\label{parameterRanges}
\end{eqnarray}

We take the 800 GeV lower limit on $M_3$ to satisfy 
the experimental lower bound on the gluino mass \cite{PDG}. 
We would like to remind our reader that the lower bound on 
the gluino mass is applicable for the CPC-MSSM and will 
change in the CPV-MSSM due to the significant modifications 
in different decay/Br's. However, from the current LHC 
results, we expect that the gluino mass bound would be in the TeV 
regime and so, to be in a conservative side, we choose 800 
GeV as the lower limit. The ranges for $\mu$, $A_t$, 
$M_{Q3}$ and $M_{U3}$ are set consulting sample test 
scans which favor relatively large values of $A_t$ and small 
values of the Higgsino mass parameter $\mu$. 
We allow the CPV phases ($\phi_3, \phi_{A_t}, 
\phi_{A_b}, \phi_{A_\tau}$) to vary between $0^o$ to $90^o$ 
independently and randomly. Other remaining parameters in this 
second scan are identical to those of the previous one (Scan 1).

\begin{figure}[!htb]\centering
\begin{tabular}{c c}
\includegraphics[angle=0,width=70mm]{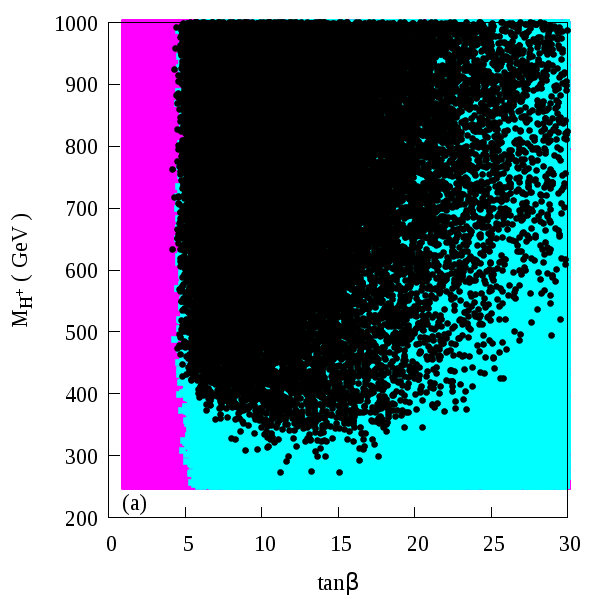} &
\hspace{1.5cm}
\includegraphics[angle=0,width=70mm]{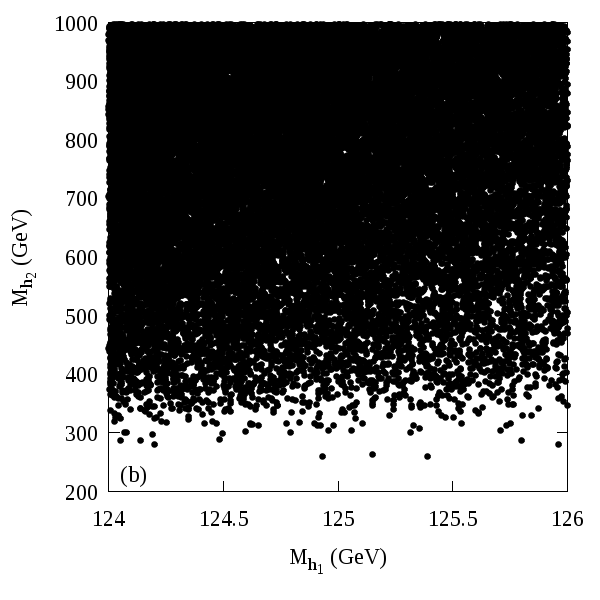} 
\end{tabular}
\vspace*{-1mm}
\caption{Constraints on the (a)
$\tan\beta - M_{H^\pm}$
and (b) $M_{h_1} - M_{h_2}$ plane respectively,
obtained after scanning the CPV-MSSM parameter space randomly,
for the input parameters mentioned in eq.(\ref{eq:B}). The
magenta/medium grey region is without any experimental 
constraint, the cyan/light grey
region is allowed only by the primary Higgs mass bound and the
black points are allowed by the set of constraints mentioned
in eq.(\ref{eq:C}) to eq.(\ref{eq:D}) and also the $d_e$
(eq.(\ref{eq:A})), $d_{Tl}$, $d_n$ and $d_{Hg}$ 
(Tab.\ref{tab:edm}) constraints.}
\label{fig:3}
\end{figure}

With these new set of CPV-MSSM parameter ranges, we 
scan the parameter space for around $10^7$ points and impose 
all the experimental constraints starting from 
eq.(\ref{eq:C}) to eq.(\ref{eq:D}) and the latest bounds on 
$d_e$ (eq.(\ref{eq:A})), $d_{Tl}$, $d_n$ and $d_{Hg}$ 
(Tab.\ref{tab:edm}), after the primary selection criterion on 
the Higgs boson mass. In Fig.\ref{fig:3}(a) and \ref{fig:3}(b), 
we present the allowed region in the $\tan\beta - M_{H^\pm}$ 
and $M_{h_1}-M_{h_2}$ planes, respectively, for the parameter 
ranges mentioned in eq.(\ref{eq:B}). Clearly, larger 
$M_{H^\pm}$ values can accommodate larger $\tan\beta$, 
possibly even going beyond 30, satisfying all the 
constraints. From Fig.\ref{fig:3}(a) we can see that
values of $\tan\beta$ smaller than around 5 are not allowed. 
In order to understand which constraint disallows $\tan\beta$
below 5, we plot the points allowed by the different set of 
constraints, where magenta/medium grey dots are the points 
without any experimental bound, the cyan/light grey dots 
are the points which obey only the Higgs mass bound and the 
black points are allowed by all the experimental constraints 
starting from eq.(\ref{eq:C}) to eq.(\ref{eq:D}) and the 
latest bounds on $d_e$ (eq.(\ref{eq:A})), $d_{Tl}$, $d_n$ 
and $d_{Hg}$ (Tab.\ref{tab:edm}). It is clear that the Higgs 
mass bound itself puts a lower limit of $\tan\beta \sim 5$. 
Here we would like to mention that, this pattern is consistent 
with the CPC-MSSM, see Ref.~\cite{Aad:2012cfr}. 
Fig.\ref{fig:3}(b) says that the second lightest Higgs 
boson (${h_2}$) can be as heavy as the ${H^\pm}$, with a 
preference for heavier masses. 


\begin{figure}[!htb]\centering
\begin{tabular}{c c}
\includegraphics[angle=0,width=70mm]{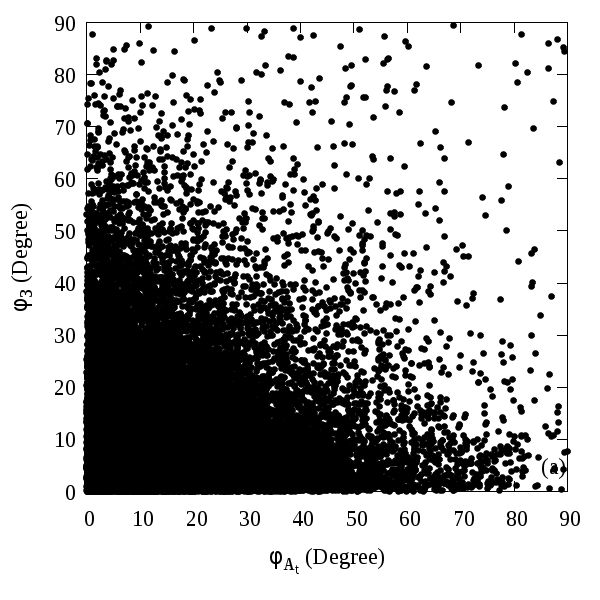} &
\hspace{1.5cm}
\includegraphics[angle=0,width=70mm]{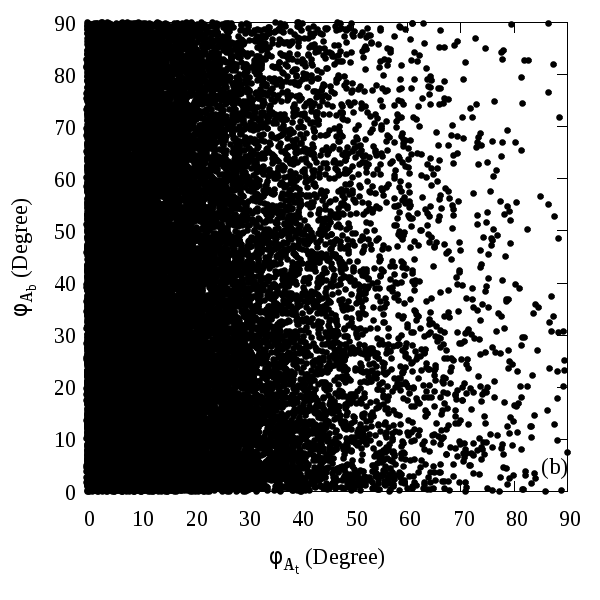}
\end{tabular}
\vspace*{-1mm}
\caption{Correlation of $\phi_{A_t}$ with (a) $\phi_3$ and 
(b) $\phi_{A_b}$. The black color code is same as in 
Fig.\ref{fig:3}.}
\label{fig:4}
\end{figure}

In Fig.\ref{fig:4} we plot the allowed points in the 
(a) $\phi_{A_t} - \phi_3$ and (b) $\phi_{A_t}-\phi_{A_b}$ 
planes. We get that most of the allowed points fall in the 
region with relatively smaller values of $\phi_{A_t}$ and 
$\phi_3$, even though there are some points with large phase 
values as well. Here we would like to note that there were no 
surviving points in the first scan (corresponding to 
eq.(\ref{eq:S})) with
$\Phi_{A_t}=\Phi_{A_b}=\Phi_{A_{\tau}}=\phi_3$ fixed at
$90^o$. In the present scan the large $M_{H^\pm}$ values
make it possible to evade the experimental constraints for
large values of the CPV phases. Fig.\ref{fig:4}(b) clearly
shows that $\phi_{A_b}$ has negligible effect. A similar
result is obtained also for $\phi_{A_\tau}$, which 
is not presented here.


\begin{figure}[!htb]\centering
\vspace*{5mm}
\begin{tabular}{c c}
\includegraphics[angle=0,width=70mm]{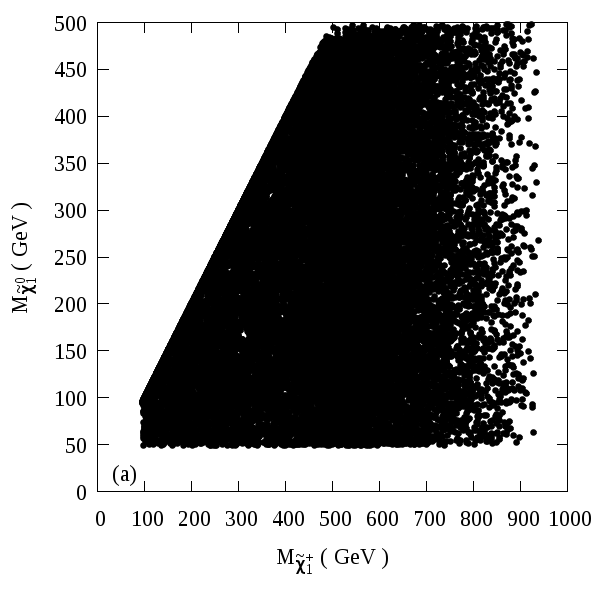}&
\hspace{1.5cm}
\includegraphics[angle=0,width=70mm]{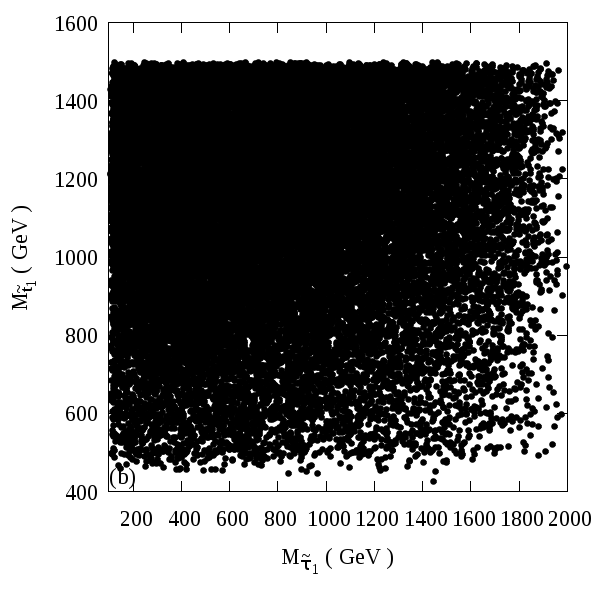}
\end{tabular}
\vspace*{-1mm}
\caption{(a) $M_{\widetilde{\chi}_1^\pm} - 
M_{\widetilde{\chi}_1^0}$  and (b) $M_{\tilde \tau_1} - 
M_{\tilde t_1}$ planes after imposing all our selection 
criteria. The black color code is same as in Fig. \ref{fig:3}.}
\label{fig:5}
\end{figure}

Fig.\ref{fig:5}(a) and (b) summarize the spectrum of
some relevant particles in the context of the CPV-MSSM
under the same conditions as previously explained, with
the same black color code as in Fig.\ref{fig:3}.
Fig.\ref{fig:5}(a) shows, in particular, that the lightest
chargino and neutralino masses could be as high as 900 GeV
and 500 GeV respectively, while Fig.\ref{fig:5}(b) says
that the lightest stop and stau masses could be as low as
450 GeV and 100 GeV respectively satisfying all the 
present experimental bounds.


\begin{figure}[!htb]\centering
\begin{tabular}{c c}
\includegraphics[angle=0,width=70mm]{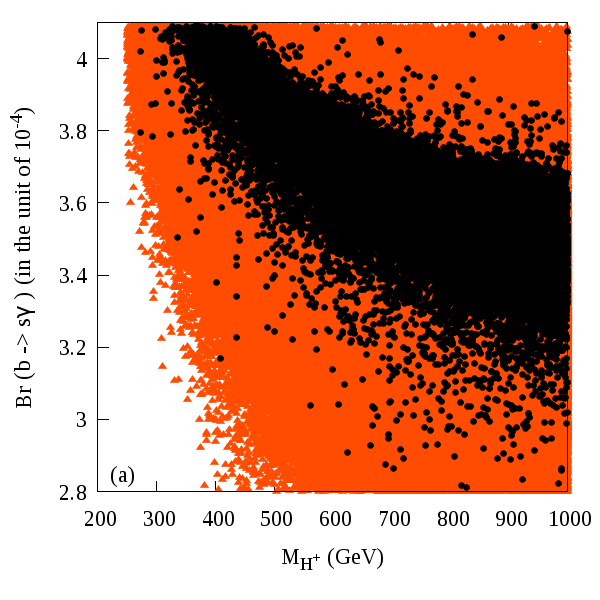} &
\hspace{1.5cm}
\includegraphics[angle=0,width=70mm]{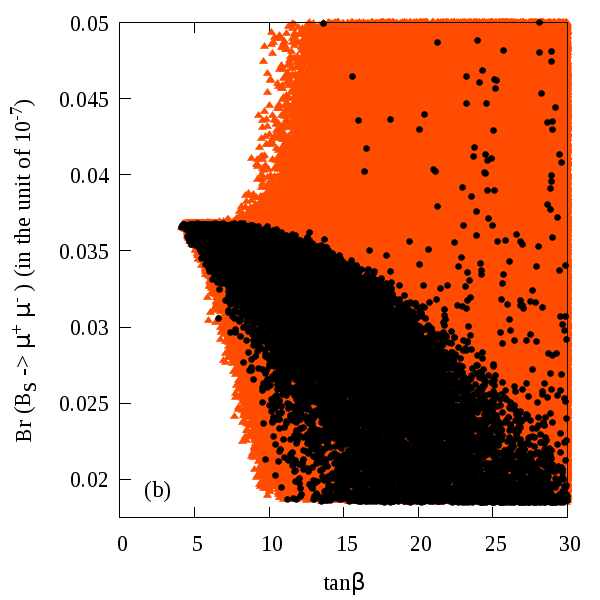}
\end{tabular}
\vspace*{-1mm}
\caption{Variation of ${Br}(b \to s \gamma)$ with the
charged Higgs mass (a) and the correlation between
${Br}(B_s \to \mu^+ \mu^-)$ and $\tan\beta$ (b). 
The black colored points satisfy all the constraints, while the 
brick red/grey points correspond to the points satisfying all 
of the constraints except the EDM ones.}
\label{fig:6}
\end{figure}

To study the constraints coming from the flavor sector, 
in Fig.\ref{fig:6}(a) we show the dependence of 
${Br}(b \to s \gamma)$ on the charged Higgs boson mass, with 
and without the imposition of the EDM constraints. Similar 
kind of correlation can be seen in Fig.\ref{fig:6}(b), where 
we plot the variation of ${Br}(B_s \to \mu^+ \mu^-)$ with 
$\tan\beta$. The black color code is the usual one which 
represents the points satisfying all the constraints, while 
the brick red/grey points satisfy all the experimental bounds 
starting from eq.(\ref{eq:C}) to eq.(\ref{eq:D}), except 
the EDM constraints, mentioned in eq.(\ref{eq:A}) and 
Tab.\ref{tab:edm}. We have already discussed in 
Sec.\ref{constraints} that significant amount of SUSY 
contribution may come from the charged Higgs loop and the 
chargino loop in $b \to s \gamma$ decay, and cancellation 
between the SUSY contribution and the SM value may occur, 
resulting in enhancement/suppression in the decay width. 
On the other hand, the SUSY contribution to the flavor 
changing $b \to s$ couplings, present in the 
${Br}(B_s \to \mu^+ \mu^-)$ decay, strongly depends on 
$\tan\beta$. From both 
Fig.\ref{fig:6}(a) and Fig.\ref{fig:6}(b), we find that the 
recent electron EDM measurements affect the CPV parameter 
space significantly, specially the region with large 
$\tan\beta$. The generic SUSY contribution to the electron 
EDM comes from charginos, neutralinos at the one loop level 
and from neutral Higgses at the two loop level. One can 
significantly reduce the one loop contributions by making the 
first two generations of sfermions very heavy 
(around 5 - 10 TeV). However, the Higgsino contribution is 
always present and it strongly depends on the Higgs boson 
coupling with the down type fermions and the coupling grows
with $\tan\beta$. So, we find that the imposition of the 
current $d_e$ bound strongly discards the large 
$\tan\beta$ regime and thereby affect both the rare 
b-decays $b \to s \gamma$ and $B_s \to \mu^+ \mu^-$
significantly. Apart from the $\tan\beta$ effect, 
non-trivial effects coming from the different loops 
associated with the different SUSY particles also play 
crucial role in accepting/discarding the parameter space 
points. 


\begin{figure}[!htb]\centering
\begin{tabular}{c c}
\includegraphics[angle=0,width=70mm]{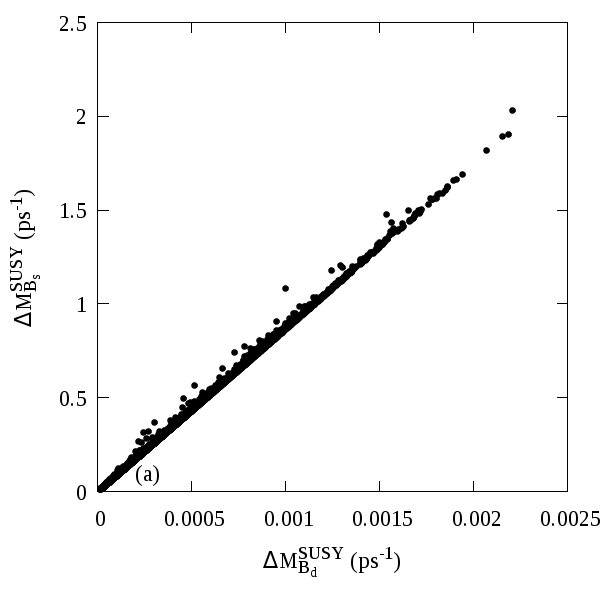}&
\hspace*{1.5cm}
\includegraphics[angle=0,width=70mm]{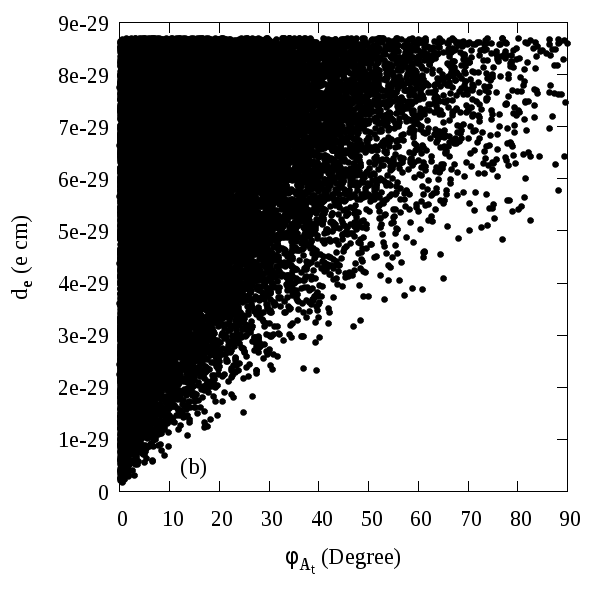}
\end{tabular}
\vspace*{-1mm}
\caption{The left panel displays the correlation in the 
$\Delta M_{B_d}^{\rm SUSY}$ - $\Delta M_{B_s}^{\rm SUSY}$  
plane, while the right panel shows the impact of the 
CPV phase $\phi_{A_t}$ on the electron EDM. 
The color code is same as in Fig.\ref{fig:3}.}
\label{fig:7}
\end{figure}


In Fig.\ref{fig:7}(a), we show the correlation in the 
$\Delta M_{B_d}^{\rm SUSY}$ - $\Delta M_{B_s}^{\rm SUSY}$ plane
where these quantities measure the SUSY contributions 
to the $B_d^0-\bar B_d^0$ and $B_s^0-\bar B_s^0$
mass differences, respectively. From the figure, it is clear 
that all the points, which survive the Higgs mass cut and the 
low energy flavor data, are well within the experimental 
$3\sigma$ limit. This figure also justifies our choice 
of neglecting the $\Delta M_{B_s}^{\rm SUSY}$ cut as a 
selection criterion. However, with reduced theoretical 
uncertainty, this constraint will play a significant role
in the CPV-MSSM parameter space. Finally, in Fig.\ref{fig:7}(b)
we show the variation of the electron EDM with the phase 
of $A_t$, namely $\phi_{A_t}$. We find that the current 
experimental electron EDM limit mostly favors smaller values 
of the CPV phases, though few parameter points may signal 
large phase values. 
We also find similar kind of behavior for all other EDMs 
against the relevant CPV phases. 


\section{Results for the LHC Higgs signals}
\label{lhcresults}

We shall now present the compatibility of the selected parameter 
space regions with the LHC measurements specific to the 
discovered Higgs boson resonance. As the first scan is found 
to be incompatible with the EDM bounds, we shall focus our
attention on the second scan. The results presented in this 
section are, therefore, those from the second scan, unless 
explicitly mentioned. 

At the LHC, the Higgs boson is dominantly produced via 
Gluon-Gluon Fusion (GGF), which at the lowest order 
occurs at one-loop level, with the Higgs boson subsequently 
decaying into $\gamma \gamma$ or 
$Z Z^* \rightarrow 4 \ell$ or 
$W W^* \rightarrow \ell \nu \ell \nu$\footnote{We 
neglect here the consideration of the $\tau^+\tau^-$
and $b\bar b$ decay modes from GGF, as corresponding
experimental errors are still very large.}.
{Now, the leading contribution of Higgs boson decays into 
$ZZ^{*}$ is via a tree-level process, whereas the Higgs decays 
into the di-photon final state via a one-loop process at 
leading order. This one-loop decay process potentially 
contains SUSY particles like stop, sbottom, stau, 
charginos and charged Higgs bosons in addition to the 
SM particles (top, bottom and 
charged gauge bosons). Conversely, in the GGF production mode 
only colored particles (top, bottom, stop and sbottom) 
contribute to the loop.} {Assuming the Narrow Width 
Approximation (NWA)}\footnote{Which is justified by 
the fact that in all models considered (SM, CPC-MSSM and 
CPV-MSSM) one has that the Higgs width is always several 
orders of magnitude smaller than the Higgs mass.}
and neglecting higher order QCD corrections at production 
level\footnote{Which would induce a different finite term 
inside the $K$-factor in the SM with respect to the CPC-MSSM 
(and CPV-MSSM as well), though with differences generally too 
small to be of relevance here.}, we define the Higgs boson 
event ratios as follows:
\begin{eqnarray}\label{eq:R}
 R_{XX} &=&  {{\Gamma( h_1 \rightarrow g g )^{\rm CPV-MSSM}}\over{{\Gamma( h \rightarrow g g )^{\rm SM}}}} \times
 {{{\rm Br}( h_1 \rightarrow X X )^{\rm CPV-MSSM}}\over{{{\rm Br}( h\rightarrow X X  )^{\rm SM}}}}, 
\end{eqnarray}
where, $XX = \gamma\gamma$ or $ZZ^*$ or $WW^*$ and 
$h_1$ is the lightest Higgs boson of the CPV-MSSM, 
while in the SM case it is marked as $h$. 


\begin{figure}[!htb]\centering
\begin{tabular}{c c}
\includegraphics[angle=0,width=70mm]{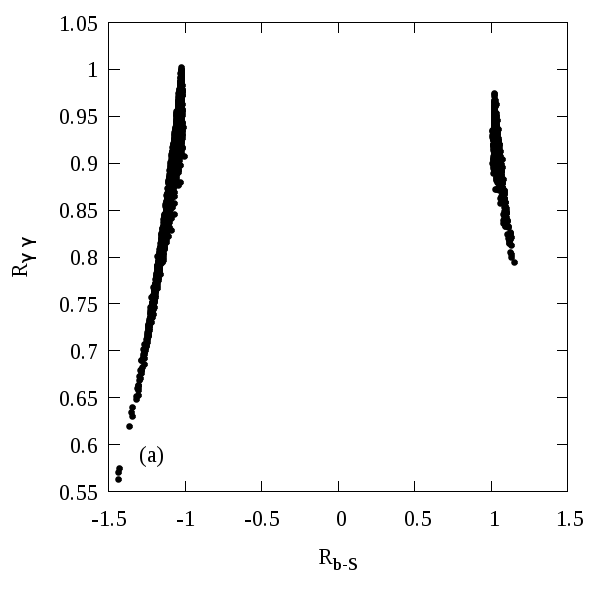}&
\hspace*{1.5cm}
\includegraphics[angle=0,width=70mm]{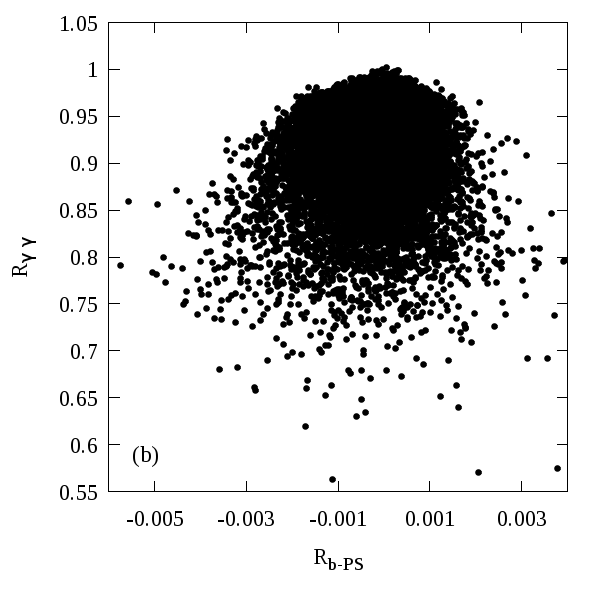}\\&\\[2.6mm]
\includegraphics[angle=0,width=70mm]{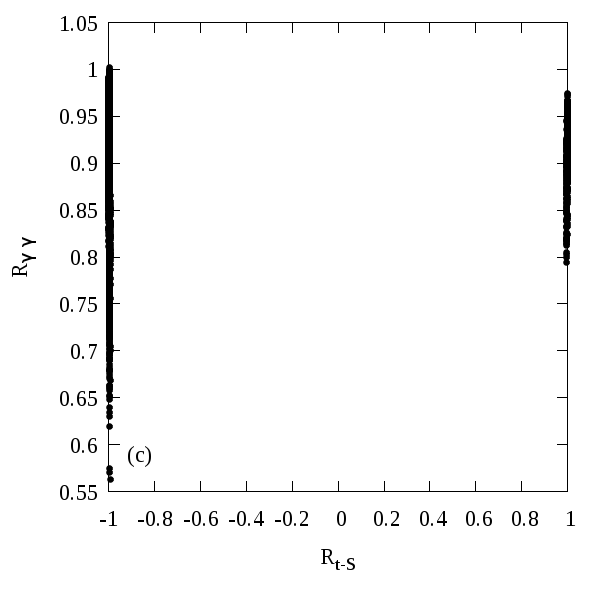} &
\hspace*{1.5cm}
\includegraphics[angle=0,width=70mm]{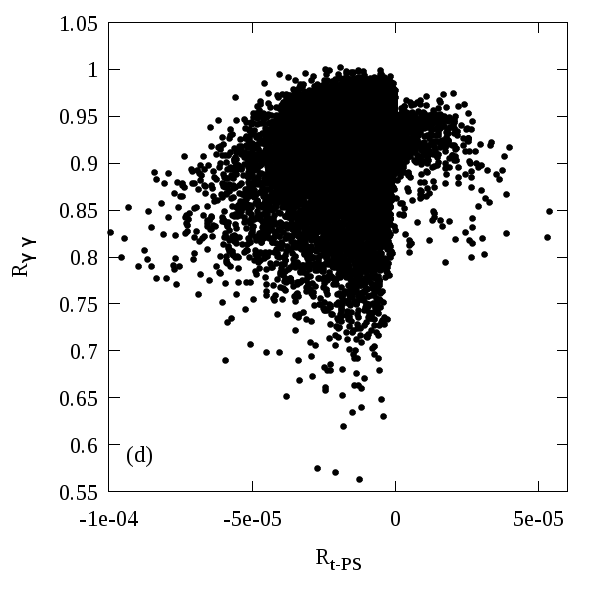}
\end{tabular}
\vspace*{-1mm}
\caption{Variation of $R_{\gamma \gamma }$ with the scalar
and pseudoscalar part of the ratio of the bottom
((a) \& (b)) and top ((c) \& (d)) quark Yukawa
couplings defined in eq.(\ref{eq:Yu}) in the CPV-MSSM.
The black color code is same as in Fig.\ref{fig:3}.}
\label{fig:8}
\end{figure}

Turning our attention to the study of effects specifically 
due to the CPV phases, notice that CPV effects enter into
eq.(\ref{eq:R}) through higher order corrections in the 
definition of the physical $h_1$ mass as well as through 
lowest order terms via the  $h_{1} \tilde{f} {\tilde{f}}^*$ 
couplings, where $\tilde f$ refers to any possible sfermion. 
In fact, as emphasized in 
Refs.~\cite{Dedes:1999zh,Dedes:1999sj,Hesselbach:2011nw,
Hesselbach:2009st,Moretti:2007th,Hesselbach:2007en,
Hesselbach:2009gw}, the most significant CPV effects are 
induced by the latter, since the former is responsible for 
mass shifts between models which are within current 
experimental uncertainties in the determination of 
the resonant Higgs mass.

In order to appreciate such specific CPV effects in our 
analysis, we find it convenient to study the ratio of 
the bottom/top Yukawa coupling for the $h_1$ state of the 
CPV-MSSM relative to SM values for the $h_1$ boson, which 
can be written as ($q=b,t$):
\begin{equation}\label{eq:Yu}
\frac{y^{\rm CPV-MSSM}_q}{y^{\rm SM}_q}  = R_{q-S} + i \gamma_5R_{q-PS}.
\end{equation}

Here $R_{q-S}$ and $R_{q-PS}$ denote the scalar and 
pseudoscalar part of the Higgs Yukawa coupling, respectively. 
The full expressions for these terms can be found in the 
CPSuperH manual \cite{cpsuperh}. When $R_{\gamma \gamma }$ 
is plotted as a function of the $b$-quark couplings, 
$R_{b-S}$ and $ R_{b-PS}$, as shown in Fig.\ref{fig:8}(a) 
and (b),  one finds that there are solutions to the 
LHC Higgs data with both positive and negative values of 
$R_{b-S}$, which is typical of the CPV-MSSM, unlike the 
case of the CPC-MSSM, which only allows for positive values. 
A similar behavior is obtained for the dependence on the 
$t$-quark couplings, as shown in  Fig.\ref{fig:8}(c) 
and (d). Further, if one recalls that the coupling of the 
top quark to the $h_1$ Higgs boson is inversely proportional 
to  sin{$\beta$} (relative to the SM case) and that we have 
varied tan{$\beta$} from 1 to 30 (which implies 
sin{$\beta$} $\sim$ 1), it is not surprising to see that 
$R_{t-S}$ remains around unity. Needless to say, 
by definition, $R_{b-PS}$ and $R_{t-PS}$ are zero in the 
CPC-MSSM, whereas both of them are non-zero in the 
CPV-MSSM, although their absolute values are much smaller 
than those for $R_{b-S}$ and $R_{t-S}$, respectively.

\begin{figure}[!htb]\centering
\begin{tabular}{c c}
\includegraphics[angle=0,width=70mm]{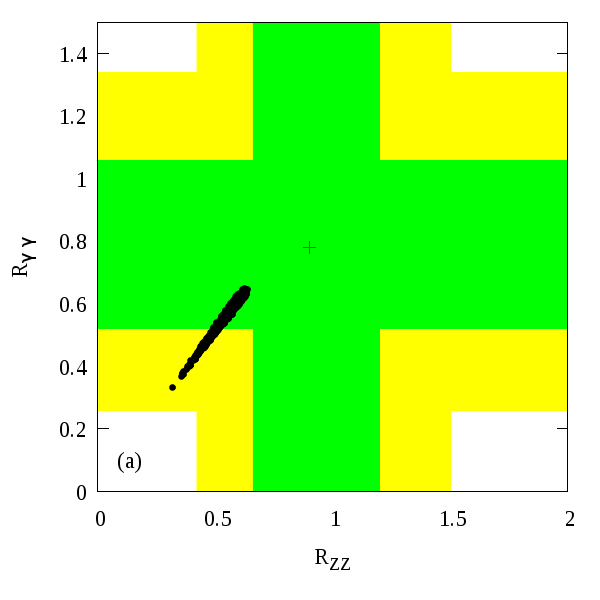} &
\hspace*{1.5cm}
\includegraphics[angle=0,width=70mm]{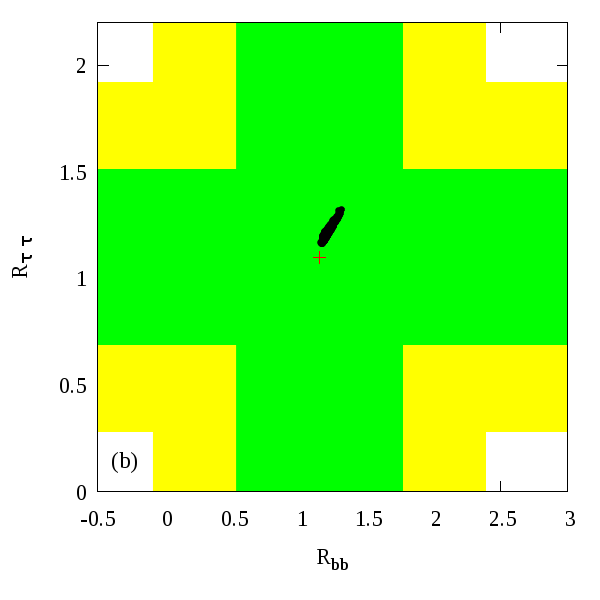}
\end{tabular}
\vspace*{-1mm}
\caption{Correlation between (a) $R_{ZZ}$ and 
$R_{\gamma\gamma}$ when the Higgs boson is produced via 
Gluon-Gluon Fusion (GGF) and (b) $R_{bb}$ and 
$R_{\tau\tau}$ when the Higgs boson is produced via 
Vector Boson Fusion (VBF), presented in comparison with 
the recent LHC data (CMS) along with the $1\sigma$ 
(green/medium grey patch) and $2\sigma$ (yellow/light 
grey patch) error bands around the experimental 
best-fit values (plus marks). The black color code is same 
as in Fig.\ref{fig:2}. Note that this scan corresponds 
to a setup in which the CPV phases are maximal i.e. $90^o$.}
\label{fig:9}
\end{figure}

We now proceed to study correlations among the  
different signal strength variables as mentioned 
above. At first, we would like to discuss the correlation 
of the signal strength variables corresponding to our 
first scan (Scan 1), although the updated results from the 
ACME EDM collaboration on the electron EDM excluded all 
the parameter space points, for the sake of completeness of the discussion. 
In Fig.\ref{fig:9}(a), we plot the correlation between 
$R_{ZZ}$ and $R_{\gamma\gamma}$. Similar 
correlation in $R_{bb}$-$R_{\tau\tau}$ plane can be seen
in Fig.\ref{fig:9}(b), when the Higgs boson is produced 
in association with a vector boson ($W/Z$). We then compare 
our results with the recent CMS 
data~\cite{cmsnewgam,cmsnewzz,cmsnewww,cmsnewcomb}.
According to the CMS collaboration, the signal strength
(our event ratios) for the $\gamma \gamma$ channel is
($0.78^{+0.28}_{-0.26}$), while for the $ZZ^*$ and $WW^*$ 
channels are ($0.9^{+0.30}_{-0.24} $) and 
($0.68 \pm 0.20$) respectively~\cite{cmsnewgam,cmsnewzz,
cmsnewww,cmsnewcomb}. We plot our results against the 
corresponding CMS results using $1\sigma$ (green/medium 
grey patch) and $2\sigma$ (yellow/light grey patch) error 
bands around the experimental best-fit values (plus marks). 
We find that the maximum value of $R_{\gamma\gamma}$ is 
$\sim$0.7 while $R_{bb}$ and $R_{\tau\tau}$ 
are always greater than 1. As our first scan is not  
compatible with the recent electron EDM bound, hence from 
now onwards, we again turn our attention to the results 
corresponding to our second scan.


\begin{figure}[!htb]\centering
\begin{tabular}{c c}
\includegraphics[angle=0,width=70mm]{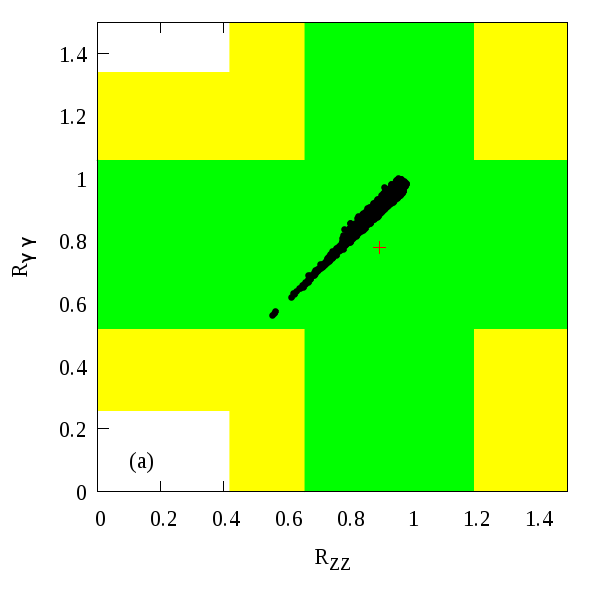} &
\hspace{1.5cm}
\includegraphics[angle=0,width=70mm]{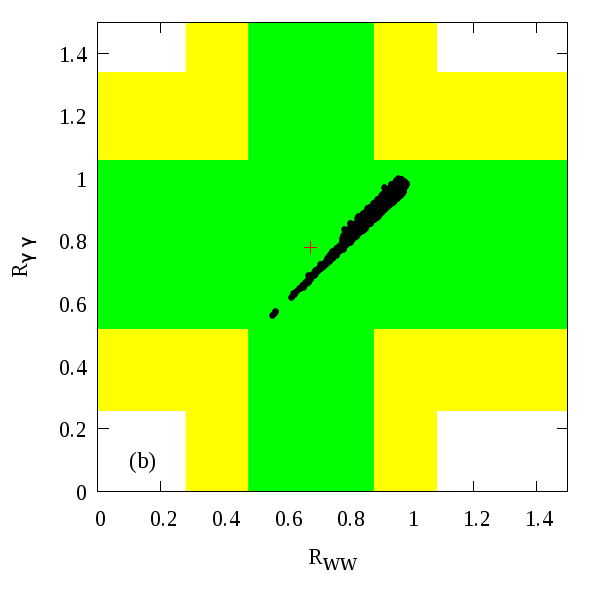}
\end{tabular}
\vspace*{-1mm}
\caption{Correlation of $R_{\gamma \gamma }$ with (a) $R_{ZZ}$
and (b) $R_{WW}$, when the Higgs boson is produced via GGF, in
comparison to the latest LHC data (CMS) along with the error
bands (the yellow/light grey and green/medium grey patches are the $2\sigma$ and $1\sigma$ uncertainty levels, respectively, 
around the experimental best-fit values, represented by the 
plus marks). The black color scheme is same as in 
Fig.\ref{fig:3}.}
\label{fig:10}
\end{figure}

We first study the correlation of $R_{\gamma \gamma }$ with 
$R_{ZZ}$ and $R_{WW}$ when the Higgs is produced via GGF 
channel. It is evident from Fig.\ref{fig:10} that these 
values are in good agreement (within 1$\sigma$) with the 
latest Higgs data as obtained by the CMS 
collaboration~\cite{cmsnewgam,cmsnewzz,cmsnewcomb} for the
$h_1 \to \gamma \gamma$, $h_1 \to ZZ^{*}$ and 
$h_1 \to WW^{*}$ channels. The observation of 
the $h_1\rightarrow  b \bar{b}$ and $h_1\to\tau^+\tau^-$
decays using GGF production mode are considered 
nearly impossible due to overshadow of QCD di-jet events. 
Hence, to discuss signal strength variables
associated with the bottoms and taus, we assume the Higgs 
production in association with a vector boson ($W/Z$) and 
the gauge bosons decaying leptonically with the Higgs 
boson decaying to a pair of $b$-jets, as both CMS and ATLAS 
have some sensitivity in this channel 
\cite{cmsnewbb,cmsnewtautau,atlasnewbb,atlasnewtautau}.
In this case, the definition of 
the corresponding event ratios will be modified to  
(here, $V = W/Z$):
\begin{eqnarray}\label{eq:RR}
 R_{YY} &=&  {{\Gamma( h_1 \rightarrow V V )^{\rm CPV-MSSM}}\over{{\Gamma( h \rightarrow V V )^{\rm SM}}}} 
 {{{\rm Br}( h_1 \rightarrow Y Y  )^{\rm CPV-MSSM}}\over{{{\rm Br}( h \rightarrow Y Y   )^{\rm SM}}}}, 
\end{eqnarray}
where $YY = b\bar b,~ \tau^+ \tau^-$.


\begin{figure}[!htb]\centering
\begin{tabular}{c c}
\includegraphics[angle=0,width=70mm]{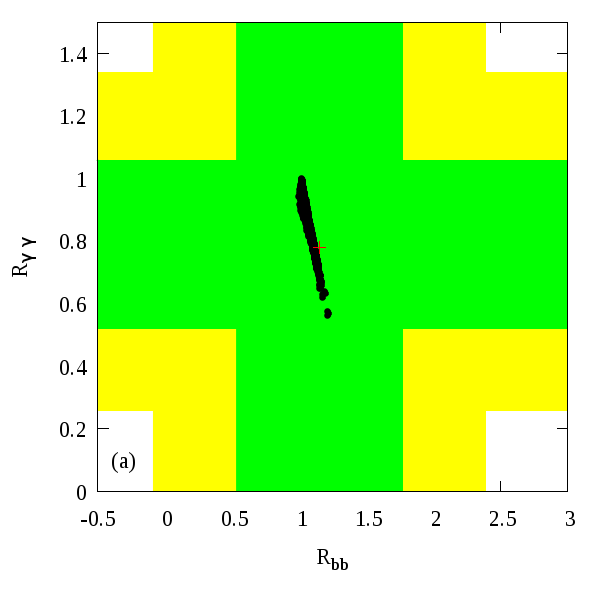} &
\hspace{15mm}
\includegraphics[angle=0,width=70mm]{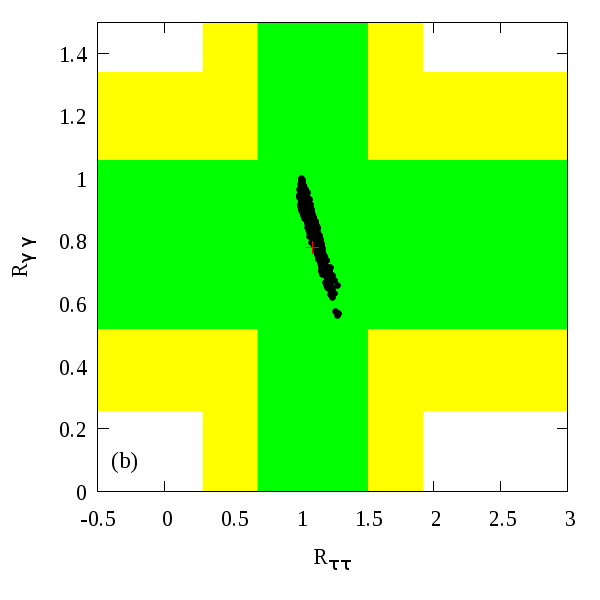}
\end{tabular}
\vspace*{-1mm}
\caption{(a) $R_{bb}$ vs $R_{\gamma \gamma}$ and 
(b) $R_{\tau \tau }$ vs $R_{\gamma \gamma }$ with the 
best-fit corresponding CMS values (plus marks). The 
yellow/light grey and green/medium grey patches are the 
$2\sigma$ and $1\sigma$ uncertainty bands, respectively. 
The black color scheme is same as in Fig.\ref{fig:3}.}
\label{fig:11}
\end{figure}


The CMS collaboration results on these decay channels are
$R_{bb} = 1.15 \pm 0.62$ and $R_{\tau \tau} = 1.10 \pm 0.41$
\cite{cmsnewcomb}, respectively. In Fig.\ref{fig:11} we 
display the scatter plots in the (a) the 
$R_{bb}- R_{\gamma \gamma}$ plane and 
(b) $R_{\tau \tau }- R_{\gamma \gamma }$ plane, with the CMS 
results. The color scheme is same as earlier. We note that 
the QCD and SUSY QCD corrections to $M_b$ ($\Delta_b$) play 
important roles in modifying the total decay width as well 
as the relevant $Br$'s of the Higgs boson
\cite{Bechtle:2012jw,Hagiwara:2012mga,Carena:2013iba}.
This primarily changes the $h_1 \rightarrow b \bar b$ 
coupling values which are reduced for large 
$\Delta_b$\footnote{Notice that $\Delta_b$ is typically 
positive for this analysis with positive $\mu$ 
\cite{Carena:2013iba}.}. In general, a reduction (enhancement) of the $h_1 \rightarrow b \bar b$ coupling 
decreases (increases) the total decay width of the Higgs 
boson. This in turn enhances (reduces) the $Br$'s to modes
like $h_1 \rightarrow \gamma  \gamma$ thereby
increasing (decreasing) $R_{\gamma \gamma}$ 
\cite{Bechtle:2012jw}. This is evident in 
Fig.\ref{fig:11}(a) that shows an anti-correlation
between the values of $R_{\gamma \gamma}$ and 
$R_{b \bar b}$. A similar kind of anti-correlation exists 
in Fig.\ref{fig:11}(b) where we show the variation in the 
$R_{\tau \tau}- R_{\gamma \gamma }$ plane. The plus marks 
represent the experimental best-fit values (CMS) for 
$R_{bb}$ and $R_{\tau \tau}$ \cite{cmsnewcomb}, with 
$1\sigma$ and $2\sigma$ error levels (green/medium grey 
and yellow/light grey patches, respectively). 

\begin{figure}[!htb]\centering
\vspace*{1cm}
\begin{tabular}{c c}
\includegraphics[angle=0,width=70mm]{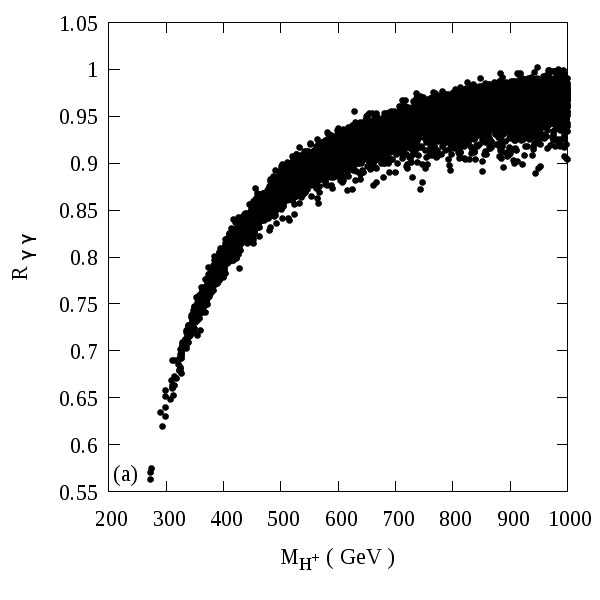}
\hspace*{1.5cm}
\includegraphics[angle=0,width=70mm]{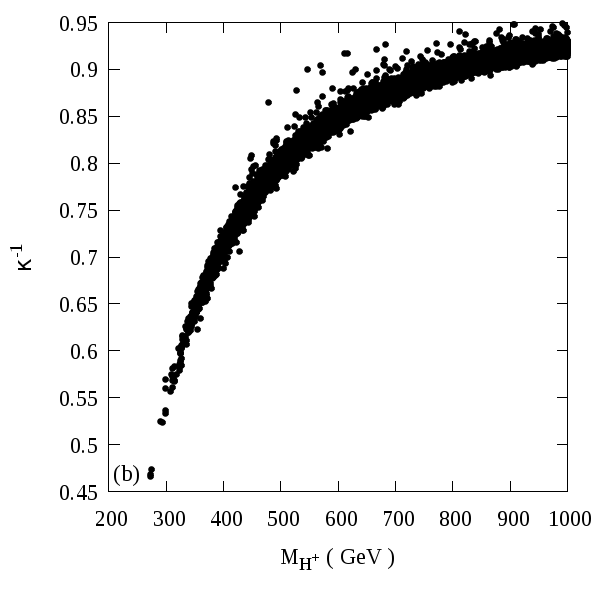}
\end{tabular}
\vspace*{-1mm}
\caption{The correlations between (a) $M_{H^\pm}$ and 
$R_{\gamma \gamma }$,
(b) $M_{H^\pm}$ and $\kappa_{bb}^{-1}$.
We define $\kappa_{bb}$ as $\kappa_{bb}\equiv\Gamma (h\to b\bar b)/\Gamma(h\to b\bar b)^{\rm SM}$. 
The black color scheme is same as in Fig.\ref{fig:3}.}
\label{fig:12}
\end{figure}

As in our second scan (Scan 2) we increase 
$M_{H^\pm}$ to 1 TeV, we find it interesting to investigate 
the role of the charged Higgs mass in the Higgs to di-photon 
decay. The charged Higgs boson contribution to the di-photon 
amplitude is usually negligible compared to the fermion and 
gauge boson loop contributions\cite{Hemeda:2013hha}. 
In Fig.\ref{fig:12}(a) we show the variation of 
$R_{\gamma \gamma}$ with $M_{H^\pm}$ and find that the 
charged Higgs contribution increases the $R_{\gamma \gamma}$ 
value. However, we find that the key role of the $H^\pm$ mass 
in $R_{\gamma \gamma}$ does not come from the $h\gamma\gamma$ 
coupling, rather it comes from the modification in the total 
decay width of the Higgs and the $hb\bar b$ coupling. To 
understand this behavior, we define a quantity $\kappa_{bb}$ 
as 
$\kappa_{bb}\equiv\Gamma (h\to b\bar b)/\Gamma(h\to b\bar b)^{\rm SM}$
keeping in mind that 
$\Gamma_{\rm tot}\approx\Gamma (h\to b\bar b)$.
In Fig.\ref{fig:12}(b), we plot $\kappa_{bb}^{-1}$ with 
$M_{H^\pm}$ and find a strong correlation between them. 
We observe that $\kappa_{bb}$ is large for smaller values of 
$M_{H^\pm}$ and it decreases with the increase of $M_{H^\pm}$. 
Since $\kappa_{bb}$ is proportional to 
$\Gamma (h\to b\bar b) \approx \Gamma_{\rm tot}$, hence 
decrease in $\kappa_{bb}$ implies decrease in total decay 
width, which in turn implies enhancement in  
Higgs to di-photon branching ratio. At the tree level, the 
charged and pseudoscalar Higgs boson masses are related as 
$m_{H^\pm}^2 = m_A^2 +m_{W^\pm}^2$. In the CPC-MSSM, the 
$hb\bar b$ coupling, at tree level, goes as 
$\sin\alpha/\cos\beta$, where $\alpha$ and $\beta$ angles 
are related as:
\begin{equation}
\tan 2 \alpha =\tan 2 \beta ~\frac{m_{H^\pm}^2-m_{W^\pm}^2 + m_Z^2}{ m_{H^\pm}^2-m_{W^\pm}^2 - m_Z^2}.
\end{equation}
From the above relation, it is obvious that the mixing angle
$\alpha$ is also dependent on $M_{H^\pm}$ and any change
in this parameter can lead to a change in $\kappa_{bb}$,
which in turn modify the $\Gamma (h\to b\bar b)$ and also
$\Gamma_{\rm tot}$.

\begin{figure}[!htb]\centering
\vspace*{10mm}
\begin{tabular}{c c}
\includegraphics[angle=0,width=70mm]{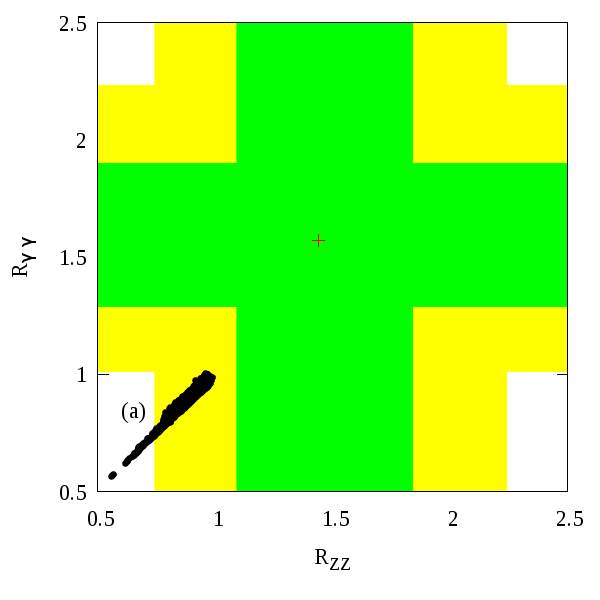} &
\hspace{1.5cm}
\includegraphics[angle=0,width=70mm]{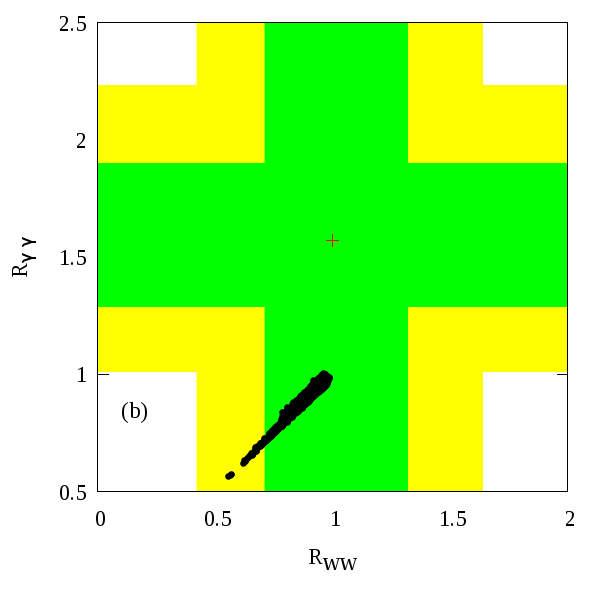}
\end{tabular}
\vspace*{-1mm}
\caption{Results on $R_{\gamma \gamma}$, $R_{ZZ}$ and 
$R_{WW}$, presented with the corresponding ATLAS results in 
(a) $R_{ZZ}- R_{\gamma \gamma}$ and 
(b) $R_{WW}- R_{\gamma \gamma}$ planes. The plus marks are 
the best-fit experimental values. The green/medium grey 
and yellow/light grey patches are 1$\sigma$ and 2$\sigma$ 
uncertainty levels, and the color scheme is same as before.}
\label{fig:13}
\end{figure}


\begin{figure}[!htb]\centering
\begin{tabular}{c c}
\includegraphics[angle=0,width=70mm]{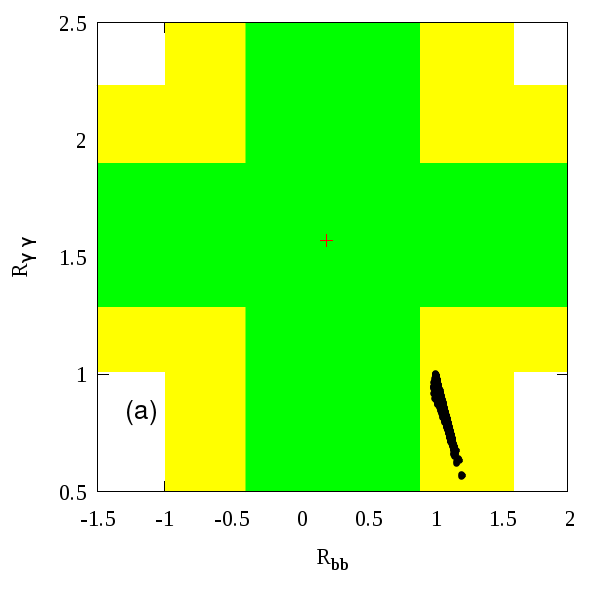} &
\hspace{1.5cm}
\includegraphics[angle=0,width=70mm]{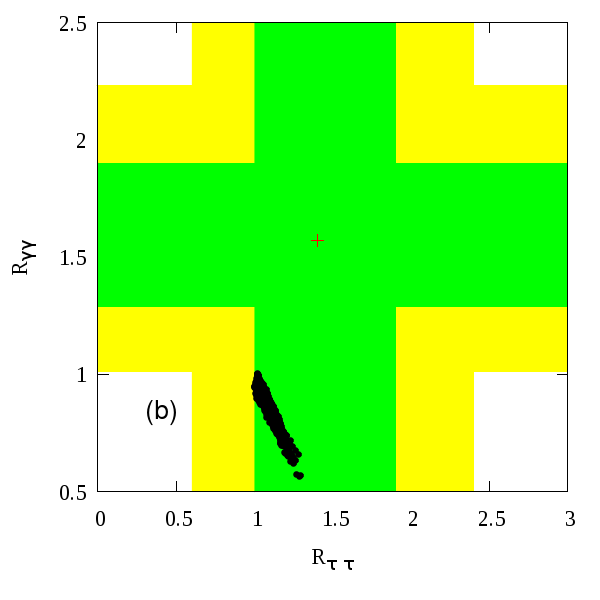}
\end{tabular}
\vspace*{-1mm}
\caption{Results on $R_{bb}$ and $R_{\tau \tau}$
presented with the corresponding ATLAS results
in (a) $R_{bb}- R_{\gamma \gamma}$
and (b) $R_{\tau \tau}- R_{\gamma \gamma}$ plane.
The green/medium grey and yellow/light grey patches are
1$\sigma$ and 2$\sigma$ error bands and the
plus marks represent the best-fit ATLAS values
for $R_{bb}$, $R_{\tau \tau}$ and $R_{\gamma \gamma}$.}
\label{fig:14}
\end{figure}

The updated ATLAS results for the signal strengths of the
aforementioned channels are:
$R_{\gamma \gamma}$ = $1.57^{+0.33}_{-0.28}$,
$R_{ZZ}$ = $1.44^{+0.40}_{-0.35}$,
$R_{WW}$ = $1.00^{+0.32}_{-0.29}$,
$R_{bb}$ = $0.2^{+0.7}_{-0.6}$ and
$R_{\tau \tau}$ = $1.4^{+0.5}_{-0.4}$~\cite{atlasnewcomb}.
We present and compare our results with those from ATLAS
in Fig.\ref{fig:13} and \ref{fig:14}. From the plots it is
clear that our $R_{\gamma \gamma}$ is just reaching the lower part of
the 2$\sigma$ region, $R_{ZZ}$ and $R_{WW}$ are almost within
the 2$\sigma$ and 1$\sigma$ bands, whereas $R_{bb}$ and
$R_{\tau \tau}$ are within the 2$\sigma$ and
1$\sigma$ error bands respectively, about the best-fit
 experimental values from the ATLAS.
However, note that there are indeed significant
discrepancies between the CMS and ATLAS results and our
results are well consistent with those from the CMS
collaboration.

\vskip 0.4cm

Before we end this section, we would like to comment on the 
phenomenological implications of the presence of non-zero pseudoscalar 
Higgs yukawa couplings. Note that, the possibility of being a 
pure CP-odd state for the observed Higgs particle has now been mostly 
ruled out \cite{Chatrchyan:2013mxa,Aad:2013wqa}, however the option of being a mixed 
CP state is still an open issue \cite{Freitas:2012kw,Belanger:2012gc,Cheung:2013kla,Cheung:2014noa}. 
A non-zero pseudoscalar Higgs yukawa coupling would affect 
several production and decay modes of the 
observed Higgs boson. For example, the gluon fusion 
process crucially depends on the Higgs couplings with the top 
and bottom quarks, while the decay of the Higgs to a pair 
of photons mostly involves the top quark coupling. A global 
analysis involving all the Higgs couplings and the 
available current LHC results 
have been performed with and without the 
current EDM constraints \cite{Cheung:2013kla,Cheung:2014noa,Cheung:2014oaa}. According 
to Ref.~\cite{Cheung:2013kla,Cheung:2014noa}, the current data 
cannot rule out the possibility of non-zero 
pseudoscalar Higgs couplings, infact it gives equally 
good fits compared to the CPC case. 
However, when current EDM bounds are considered, 
the Higgs pseudoscalar couplings are restricted 
to approximately $10^{-2}$ \cite{Cheung:2014oaa}. Note that, from 
Fig.~\ref{fig:8}(b) and (d), it is evident that 
our findings are in good agreement with the 
results obtained from a dedicated global fit.
The measurement of the CP properties 
of the Higgs boson at the LHC mostly rely on its couplings 
to massive vector bosons. It has been shown in Ref.~\cite{Dolan:2014upa} that 
the gluon fusion process could be sensitive enough 
at the 14 TeV run of the LHC to study the CP properties of 
the Higgs boson. In fact, the presence of non-trivial CPV
Higgs couplings would have important implications in the electroweak 
baryogenesis \cite{Shu:2013uua}.


\section{Conclusions}
\label{conclusions}

The discovery of a Higgs-like resonance with a mass close to 
$125$ GeV by both the multi-purpose experimental 
collaborations ATLAS and CMS operating at the LHC 
created a great interest in understanding the ultimate 
means adopted by nature for mass generation. The 
precise determination of its spin, CP properties and couplings 
to the SM fermions and gauge bosons are highly 
crucial to know the exact dynamics of electro-weak 
symmetry breaking. Although the measurements done so far indicate 
that this Higgs-like boson is compatible with the 
SM hypothesis, however due to large uncertainties in 
some of the Higgs detection channels, 
one still has the possibility of testing this object 
as being a candidate of some BSM physics. 

With this motivation in mind, we scan the CPV-MSSM 
parameter space in order to accommodate the 125 GeV Higgs
boson with signal event rates consistent with the 
observed LHC data and all other available experimental 
bounds till date. It is well known that any new 
source of CP violation (above and 
beyond what is embedded in the SM) would lead to additional
contributions to the various EDMs. Therefore, while scanning 
the CPV-MSSM parameter space, we also enforce different EDM 
constraints, namely the electron, neutron, Mercury 
and Thallium EDMs. In addition, we vary the CPV phases 
of the gaugino mass parameter $M_3$, trilinear couplings 
$A_t, A_b $ and $A_\tau$ from $0^o$ to $90^o$. 
Note that we set other CPV phases like $\phi_1$ 
(phase of $M_1$),
$\phi_2$ (phase of $M_2$), phases of the first two 
generations of fermions to zero since these variables
affect the Higgs sector negligibly. We further impose 
several low energy constraints, mainly coming from the 
different heavy flavor physics processes. 

We perform two separate parameter space scans: 
(a) with some of the CPV phases to their 
maximal value ($90^o$) and (b) varying these 
phases from $0^o$ to $90^o$. 
For both these two scans, other parameters vary 
randomly within some specified ranges. We see that maximal 
phase scenario (case (a)) is ruled out by the current EDM 
measurements, specially updated electron EDM measurement. 
However, we find significant amount of parameter space 
points, in case (b), satisfying all the constraints 
including EDMs. As expected, we see that 
relatively smaller values (c.f. Fig.\ref{fig:7}(b)) of these 
CPV phases are favored by the EDM constraints. We also 
calculate the signal rates of the Higgs boson in 
the $gg \to h_1 \to \gamma \gamma $, 
$ gg \to h_1 \to ZZ^{\ast} \to 4\ell $, 
$ gg \to h_1 \to WW^{\ast} \to \ell \nu \ell \nu$, 
$pp \rightarrow V h_1 \rightarrow V b\bar b$ and
$pp \rightarrow V h_1 \rightarrow V \tau^+\tau^-$
($V \equiv W^\pm, Z$)
channels and find that over a large expanse of parameter space of
CPV-MSSM, our results are compatible (within 1$\sigma$) 
with the observed data from the CMS collaboration, while most of them are 
still consistent within $2\sigma$ of the ATLAS 
results. 

Although, our results do not differ significantly
from those of the CPC-MSSM, which are available
in the literature. However, we find some 
interesting results in terms of
some of the observables of CPV-MSSM. The couplings of the 
Higgs boson with the bottom quark and top quark are very 
important to claim the discovery of the observed particle 
as the SM Higgs boson. We find that the imaginary part of 
the top and bottom Yukawa couplings can take very small but 
non-zero values even after satisfying all the recent 
updates from both the CMS and ATLAS collaborations 
(in terms of the signal strength variable $\mu$, or R 
in our case) within $1-2\sigma$ uncertainty. Moreover, we 
also find an interesting result from the correlation plots 
of the different signal strength variables. We do not find 
any significant excess in the di-photon decay 
mode (in both scan 1 and scan 2), but we do see excess of
events over the SM predictions for both the $b\bar{b}$
and $\tau^{+}\tau^{-}$ decay modes, i.e., in $R_{bb}$ and 
$R_{\tau\tau}$, when the Higgs boson is produced in 
association with the SM gauge bosons $W$ or $Z$. The 
suppression in the di-photon decay mode with simultaneous 
enhancement in $b\bar{b}$ and $\tau^{+}\tau^{-}$ decay 
modes with respect to the SM prediction and presence of 
non-zero imaginary parts of the top and bottom Yukawa 
couplings, could be an interesting signature of this model. 
We briefly discuss the phenomenological implications of the 
presence of such non-zero pseudoscalar Higgs coupling. 
In addition to that, we also find that it is possible to 
have a Higgs mass of about 125 GeV with relatively small 
$\tan\beta$, large $A_t$ and a light stop, which is consistent with 
the current supersymmetric searches at the LHC.

Altogether then, these findings point to the fact 
that the CPV-MSSM provides an equally competitive solution 
(like its CPC counterpart) to the updated LHC 
Higgs data, in fact offering very little in the way of 
distinction between these two SUSY models (CPC-MSSM and 
CPV-MSSM) at the current LHC run. Improvement in different 
Higgs coupling measurements is necessary in order to test 
the possibility of probing the mild dependence of these 
CPV phases in the Higgs sector of the minimal SUSY 
realization.


\section*{Acknowledgments}

A.C. would like to acknowledge the hospitality provided 
by IIT Guwahati where part of this work was done. 
J.L.D.-C. acknowledges support from VIEP-BUAP and 
CONACYT-SNI (Mexico). The work of S.M. is supported 
in part through the NExT Institute. The work of P.P. and 
B.D. is supported by a SERC, DST (India) project, 
SR/S2/HEP-41/2009. A.C. would like to thank Dr. Debottam Das 
and Dr. Biplob Bhattacherjee for useful discussions.


\bibliographystyle{unsrt}

\end{document}